# Plasmonic nanocrystals with complex shapes for photocatalysis and growth: Contrasting anisotropic hot-electron generation with the photothermal effect

*Artur Movsesyan, Eva Yazmin Santiago, Sven Burger, Miguel A. Correa-Duarte, Lucas V. Besteiro,[*] Zhiming Wang,[*] and Alexander O. Govorov[*]*

A. Movsesyan, Z. Wang,
Institute of Fundamental and Frontier Sciences, University of Electronic Science and Technology of China, Chengdu 610054, China
E-mail: zhmwang@uestc.edu.cn
A. Movsesyan, E. Y. Santiago, A. O. Govorov,
Department of Physics and Astronomy and Nanoscale and Quantum Phenomena Institute, Ohio University, Athens OH 45701, USA
E-mail: govorov@ohio.edu
S. Burger
Zuse Institute Berlin, 14195 Berlin, Germany; JCMwave GmbH, 14050 Berlin, Germany
M. A. Correa-Duarte, L.V. Besteiro
CINBIO, Universidade de Vigo, 36310 Vigo, Spain
E-mail: lucas.v.besteiro@uvigo.es
Z. Wang
Institute for Advanced Study, Chengdu University, Chengdu 610106, China

**Abstract**: In plasmonics, and particularly in plasmonic photochemistry, the effect of hot-electron generation is an exciting phenomenon driving new fundamental and applied research. However, obtaining a microscopic description of the hot-electron states represents a challenging problem, limiting our capability to design efficient nanoantennas exploiting these excited carriers. This paper addresses this limitation and studies the spatial distributions of the photophysical dynamic parameters controlling the local surface photochemistry on a plasmonic nanocrystal. We found that the generation of energetic electrons and holes in small plasmonic nanocrystals with complex shapes is strongly position-dependent and anisotropic, whereas the phototemperature across the nanocrystal surface is nearly uniform. Our formalism includes three mechanisms for the generation of excited carriers: the Drude process, the surface-assisted generation of hot-electrons in the sp-band, and the excitation of interband d-holes. Our computations show that the hot-carrier generation originating from these mechanisms reflects the internal structure of hot spots in nanocrystals with complex shapes. The injection of energetic carriers and increased surface phototemperature are driving forces for photocatalytic and photo-growth processes on the surface of plasmonic nanostructures. Therefore, developing a consistent microscopic theory of such processes is necessary for designing efficient nanoantennas for photocatalytic applications.

## 1. Introduction.

The free electrons in the conduction band of a metal collectively oscillate when excited with an external electromagnetic field. By nanostructuring these metals, we can induce strong resonant oscillations that couple strongly with radiation.[1–3] These modes, known as localized surface plasmon resonances, appear in the optical frequency range for noble metal nanostructures of sub-wavelength dimensions. Importantly, such metal nano-objects, due to their plasmon resonances, enable a spatial and temporal manipulation of light at the nanoscale.[4–12] Metal nanomaterials and nanocrystals (**NC**s) can confine and concentrate electromagnetic fields near their surface through the excitation of LSPRs. There are two main optical processes associated with a metal NC: scattering of photons, which can be observed in the far-field zone, and light absorption, which localizes the electromagnetic inside a NC. The optical excitation generates a non-equilibrium distribution of electrons in the metal NC, which involves both low-energy intraband carriers (Drude currents) and hot carriers with high energies.[9,13–16] For the latter, there are two excitation pathways: (1) the generation of hot electrons (**HE**s) inside the sp-band via surface scattering and (2) the creation of energetic holes in the d-band through interband transitions.[9,17–22]Numerous theoretical and experimental studies have contributed to a better understanding of the generation of hot carriers in plasmonic nanostructures and their injection to other media.[7,8,12,14,22–26] Regarding the process of photo-excitation in a NC, the distribution of hot-carriers excited above and below the Fermi energy of the metal depends on the incoming photon energy, the metal band structure, and the relaxation pathways available to the system.[9,13,15,27] First, the non-thermalized carriers (electrons and holes), which we label here as "hot", experience ultrafast (on the fs-scale) non-radiative relaxation in a metal NC through electron-electron scattering.[7,23] Afterwards, the distribution of excited carriers reaches a steady state balancing the continuous optical excitation and electron-phonon collisions (on the picosecond scale). This process results in the energy transfer to the atomic lattice and leads into an increase of the lattice temperature of the metal. It can also be regarded as photoheating.

A particularly active research direction within the field of HE physics is its application in the context of plasmonic photochemistry. Within this area of study, we can find a variety of heterogeneous photochemical regimes and effects (Figure 1). Photochemical transformations can occur in the bulk of a liquid matrix[28] (Figure 1a); in this case, HEs are first generated inside a NC and then transferred to the liquid, creating reactive chemical species. Excited plasmonic carriers and their energy can also be applied in the nanofabrication of hybrid nanostructures.[29] For instance, HEs and hot-holes (**HH**s) were used for site-selective reactions leading to photo-growth of NCs with complex shapes (Figure 1b,c,j).[19,30,31] The use of chiral nano-assemblies may enable polarization-sensitive photochemical reactivity promoted by HEs (Figure 1k).[32] Unlike the surface-assisted HE process, the excitation of energetic d-holes is a bulk process and was also utilized for various photoreactions (Figure 1d,e),[17,18] including the challenging synthesis of hydrocarbons from $CO_2$ (Figure 1f).[33,34] Within a hybrid Au-$TiO_2$ system, the injection of non-thermal hot carriers excited through different mechanisms can coexist (Figure 1g),[35] so that both intraband HEs and hot d-holes in the Au component, together with the interband electron-hole pairs in the semiconductor, can contribute to a photoreaction.[35] One can map HEs and related hot spots by observing the photochemical growth of plasmonic nano-islands (Figure 1h, top graph).[36] Contrasting with hot-carrier photochemistry, a pure photo-thermal mechanism can be used to grow a semiconductor layer on gold NCs using low energy (~1.1 eV) irradiation in the IR spectral range (Figure 1h, bottom graph).[37] In plasmonic photochemistry, hot-electron effects may co-exist with photoheating, and such behavior was demonstrated in ref. [38]. In the above experimental examples, surface photochemistry rates can be either isotropic (Figure 1d-g) or strongly anisotropic (Figure 1a-c, h-k); therefore, it should be critical to fully understand and characterize the related spatial distributions of HEs generation and temperature. In the case of the intraband HEs (Figure 1a,g), the generation process occurs at the surface of a NC,[9,14] and the relevant physical quantity to compute should be the surface map of the rate of HE generation.[14] For the hot d-holes (Figure 1d-f), the photogeneration occurs in the bulk of a NC, and we consequently need to look at bulk excitation distributions, which will be computed in the next sections; such bulk distributions will in general be non-uniform inside a NC and may lead to the anisotropy of photochemical reaction rates at the surface. The photothermal chemical processes depend on the phototemperature near the NC, so that, again, a surface temperature map should be a relevant quantity to study. The above considerations define our main motivation for this paper – we aim to computationally investigate the spatial distributions of the generation rates for the HEs, d-holes, and phototemperature in NCs with complex shapes, critically discussing their different properties.

In this paper, we investigate HE and photothermal effects in plasmonic NCs with complex shapes in both substrate and solution settings. These photophysical effects are responsible for plasmon-driven surface photochemistry and play major roles in many recent experimental realizations of plasmonic photocatalysis. Importantly, we develop and use convenient computational formalisms to reveal such spatial distributions. In particular, this paper describes the shape and polarization dependences of the spatial distributions of the physical quantities mentioned above, comprehensively discussing their different properties. We evaluate these distributions for gold NCs with complex shapes, develop convenient computational formalisms for their study, and compute their efficiencies. The generation processes of our interest involve excited low-energy Drude electrons, high-energy intraband electrons, and high-energy d-holes. The generation rates of energetic carriers involve both surface and bulk mechanisms. As a parallel task, we examine the spatial distributions of the phototemperature in NCs with complex shapes

Our results reveal that the spatial distribution of the generation rates of high-energy carriers (intraband HEs and d-holes) is typically shape- and polarization-dependent. In particular, these physical magnitudes reflect the presence of hot spots in a NC,[15,39,40] can be strongly anisotropic, and depend on the experimental setting. We observe that, whereas the HEs rates are distributed non-uniformly inside NCs, the phototemperature is uniform within a NC. Such uniform phototemperature distributions originate from the high thermal conductivity of noble metals. Furthermore, our theory covers two experimentally relevant settings: the unidirectional illumination of NCs on a substrate and the solution case with randomly-oriented NCs (Figure 2a). As illustrated by Figure 1, the HE and photothermal processes are the main driving forces for local photochemical reactions in many experiments with plasmonic NCs. The goal of this paper is to present a consistent and transparent theory describing relevant photophysical and photochemical mechanisms.

.

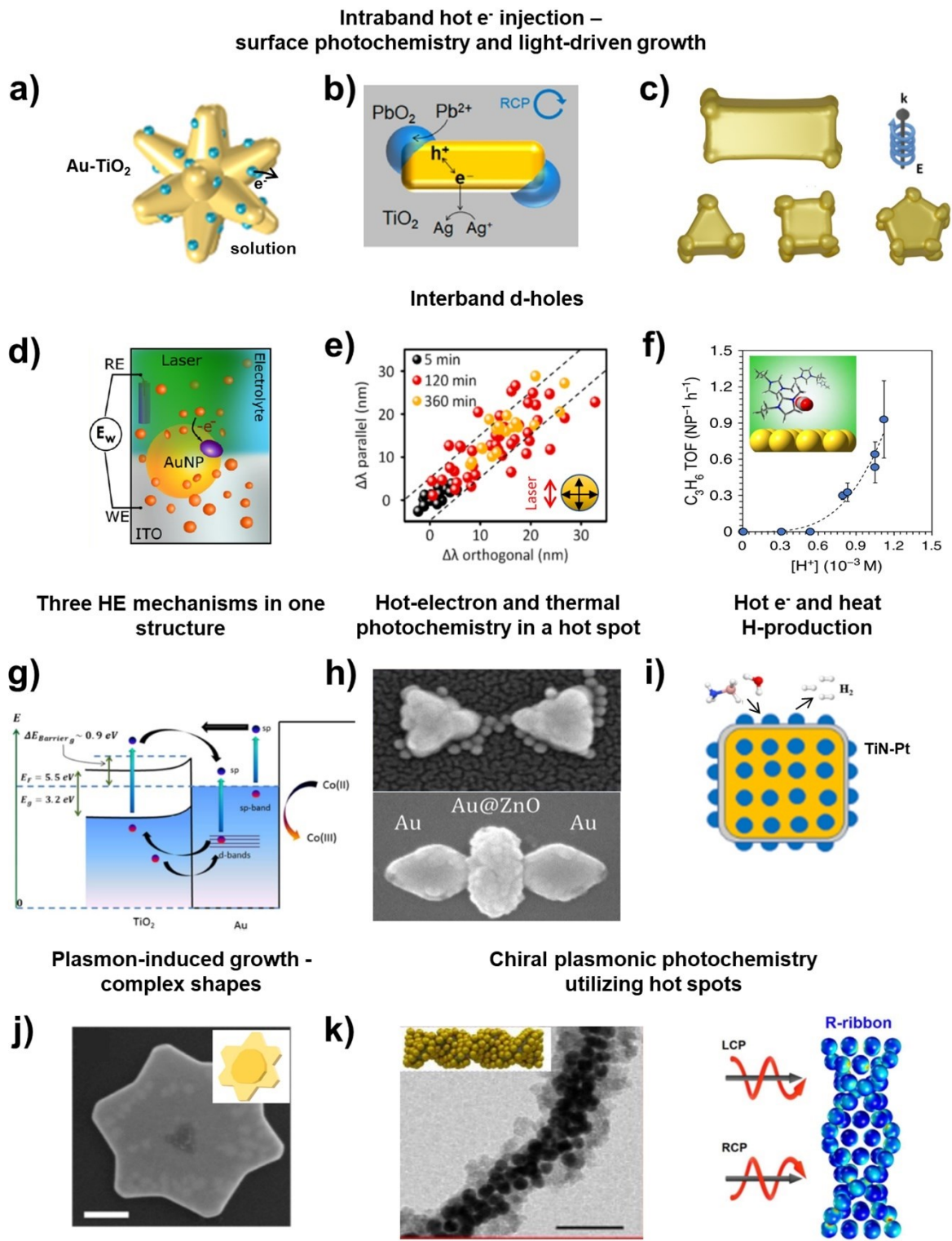

**Figure 1.** Examples of hot-carrier photochemistry. a) The comparative study has shown that a nanostar, among different shapes of nanostructures, has the highest efficiency for hot-electron surface photochemistry. Reproduced with permission.[14] Copyright 2020, American Chemical Society. The relevant experimental work on the nanostars can be found in Ref. [38]. b) The hot-

electron driven anisotropic growth of lead dioxide ($PbO_2$) on gold cuboids leads to the development of a chiral shape. Reproduced with permission.[30] Copyright 2018, American Chemical Society. c) Resulting geometries of theoretically computed anisotropic growth of gold nanoprisms under circularly polarized light. Reproduced with permission.[31] Copyright 2021, American Chemical Society. d) Photo-excited d-holes in gold nanospheres assist the process of electropolymerization of aniline molecule. Reproduced with permission.[18] Copyright 2019, American Chemical Society. e) The hot d-holes driven growth of silver shell on the gold nanosphere. The graph shows the plasmonic blue shifts of several core@shell nanoparticles measured at light polarization parallel and perpendicular to the laser polarization used during the growth. The linear behavior shows that the growth in both directions is isotropic. Reproduced with permission.[17] Copyright 2019, American Chemical Society. f) Illustration of plasmonic photosynthesis of $C_1$–$C_3$ hydrocarbons starting from carbon dioxide and with the assistance of an ionic liquid; the data show the turnover frequency (TOF) for the hydrocarbon product $C_3H_6$. Reproduced with permission.[33] Copyright 2019, Springer Nature. g) Generation of enhanced photocurrents in Au-$TiO_2$ nanocomposites. This process involves the hot electrons excited within the sp-band, the hot holes excited from the d-band, and the excitons generated via interband absorption in $TiO_2$. Reproduced with permission.[35] Copyright 2015, Springer Nature. h) The hot-carrier assisted growth of silver nanocrystals in the hot spots of a gold nano-bowtie (top graph). Reproduced with permission.[36] Copyright 2017, Springer Nature. Photothermal growth of a ZnO layer on a gold butterfly-like nanostructure (bottom graph). The growth of the ZnO layer is isotropic on the middle element (nanobar) of the nanostructure.[37] The excitation wavelength is resonant with the nanobar, therefore the growth is observed solely on it. Reproduced with permission.[37] Copyright 2019, American Chemical Society. i) This panel shows the TiN-Pt hybrid nanostructures used for enhanced photocatalysis, which involves both the collective heating effect and the hot carriers. Reproduced with permission.[38] Copyright 2020, American Chemical Society. j) Hot-hole driven anisotropic growth of gold nanostars forming from Au nanoprisms. Reproduced with permission.[19] Copyright 2020, American Chemical Society. k) Right-handed (R) ribbon assembled from gold nanospheres on a silica support for polarization-sensitive enhanced photodegradation assisted by chiral hot-spots.[32] The left graph shows the SEM and 3D image of the chiral geometry. The right graph maps the chiral hot-spots in R-ribbon. Reproduced with permission.[32] Copyright 2022, American Chemical Society.

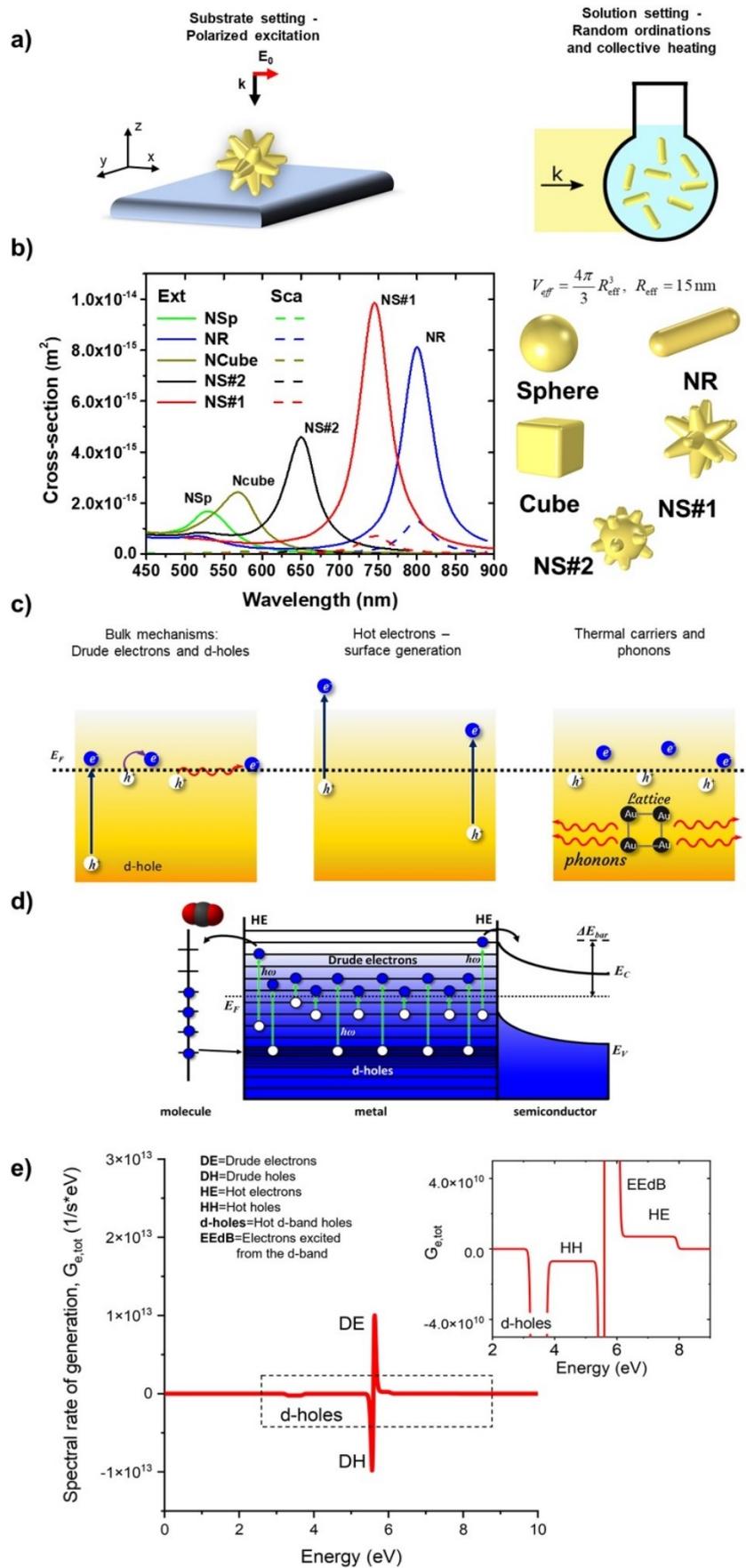

**Figure 2.** a) Models of a nanocrystal located on a substrate (substrate setting) and randomly oriented NCs (solution setting). b) Scattering and extinction spectra of differently shaped NCs;

for all NCs, the volume is fixed, and its effective radius is defined as $R_{eff} = 15\,\text{nm}$. c) Hot-electron and thermal processes in optically-driven NCs. d) Energy diagram of an optically-driven NC, including two mechanisms of transfer of excited electrons to the environment. e) As an example, this graph shows a spectral rate of generation of excited electrons in a 30-nm spherical NC, taken from Ref. [14]. Reproduced with permission.[14] Copyright 2020, American Chemical Society.

## 2. Formalisms for hot-electron generation and phototemperature.

### 2.1 Optics of anisotropic nanocrystals.

We start with the external field of a monochromatic electromagnetic wave exciting a metal NC: $\mathbf{E}_{ext} = \text{Re}\,\tilde{\mathbf{E}}_0 \cdot e^{-i\omega t} = \tilde{\mathbf{E}}_0 \cos[\omega t]$, where $\tilde{\mathbf{E}}_0$ is the real amplitude. Correspondingly, the monochromatic local electric field inside and beyond of the NC takes the following form:

$$\mathbf{E}(\mathbf{r}) = \text{Re}[\tilde{\mathbf{E}}_\omega(\mathbf{r}) \cdot e^{-i\omega t}], \qquad (1)$$

where $\mathbf{E}_\omega(\mathbf{r})$ is the complex amplitude of the electric field at position $\mathbf{r}$ and for frequency $\omega$. The optical power absorbed by a metal NC is given by the integral[41]

$$P_{abs} = \left\langle \int_{NP} dV\, \mathbf{j} \cdot \mathbf{E} \right\rangle_{time} = \text{Im}[\varepsilon_{metal}] \cdot \varepsilon_0 \frac{\omega}{2} \int_{NC} dV\, \tilde{\mathbf{E}}_\omega \cdot \tilde{\mathbf{E}}_\omega^*, \qquad (2)$$

where $\mathbf{j}$ is the current density inside the NC, and $\varepsilon_{metal}$ is the dielectric constant of the NC, defined below. The integral is taken over the NC volume. For the absorption cross-section of a NC, we have then:

$$\sigma_{abs}(\omega) = \frac{P_{abs}}{I_0},$$

where $I_0$ is the energy flux of the incident electromagnetic wave, given by the following formula:

$$I_0 = \frac{\varepsilon_0 c_0 \sqrt{\varepsilon_{env}}}{2} \cdot \tilde{E}_0^2,$$

where $c_0$ is the light speed in vacuum, $\varepsilon_{env}$ is the dielectric constant of the surrounding environment. In our numerical computation we used $\varepsilon_{env} = 2$, which is an effective dielectric constant chosen to represent a matrix media for a NC covered by small TiO$_2$ clusters and immersed in a water (Figure 1a).[28,42] The TiO$_2$-Au system from Refs. [28,42] serves for us as a model. The incident flux intensity is taken to be 400 W/cm$^2$ for all computations done. For a NC, the total extinction power, $P_{tot}$, can be split up in two terms:

$$P_{tot} = P_{scat} + P_{abs}, \quad (3)$$

where $P_{scat}$ and $P_{abs}$ are the scattered and absorbed powers, respectively. Correspondingly, the extinction cross-section takes the form:

$$\sigma_{ext}(\omega) = \sigma_{scat}(\omega) + \sigma_{abs}(\omega),$$

where $\sigma_{scat}(\omega)$ and $\sigma_{abs}(\omega)$ are the scattering and absorption cross-sections, respectively. In our study we focused on relatively small NCs, for which absorption dominates.

Reproducing typical experimental settings in photocatalysis, we consider two main excitation configurations. The first configuration involves illuminating with linearly polarized light when the nanostructure is placed on a surface, as depicted in Figure 2a. The second configuration represents a solution setting and includes NCs in water with random orientations, as shown in Figure 2a. This configuration corresponds to the case of the non-polarized excitation of a NC. Obtaining a solution spectrum involves averaging the response of the NC over six light incident configurations. For the two above settings (Figure 2a), we compute the properties of gold NCs with various shapes (Figure 2b), including nanospheres (**NSp**s), nanorods (**NR**s), nanocubes (**NCube**s), and nanostars with short (**NS#2**) and long spikes (**NS#1**). Finally, we note that we computed the electromagnetic and photothermal responses of these NCs using COMSOL Multiphysics.

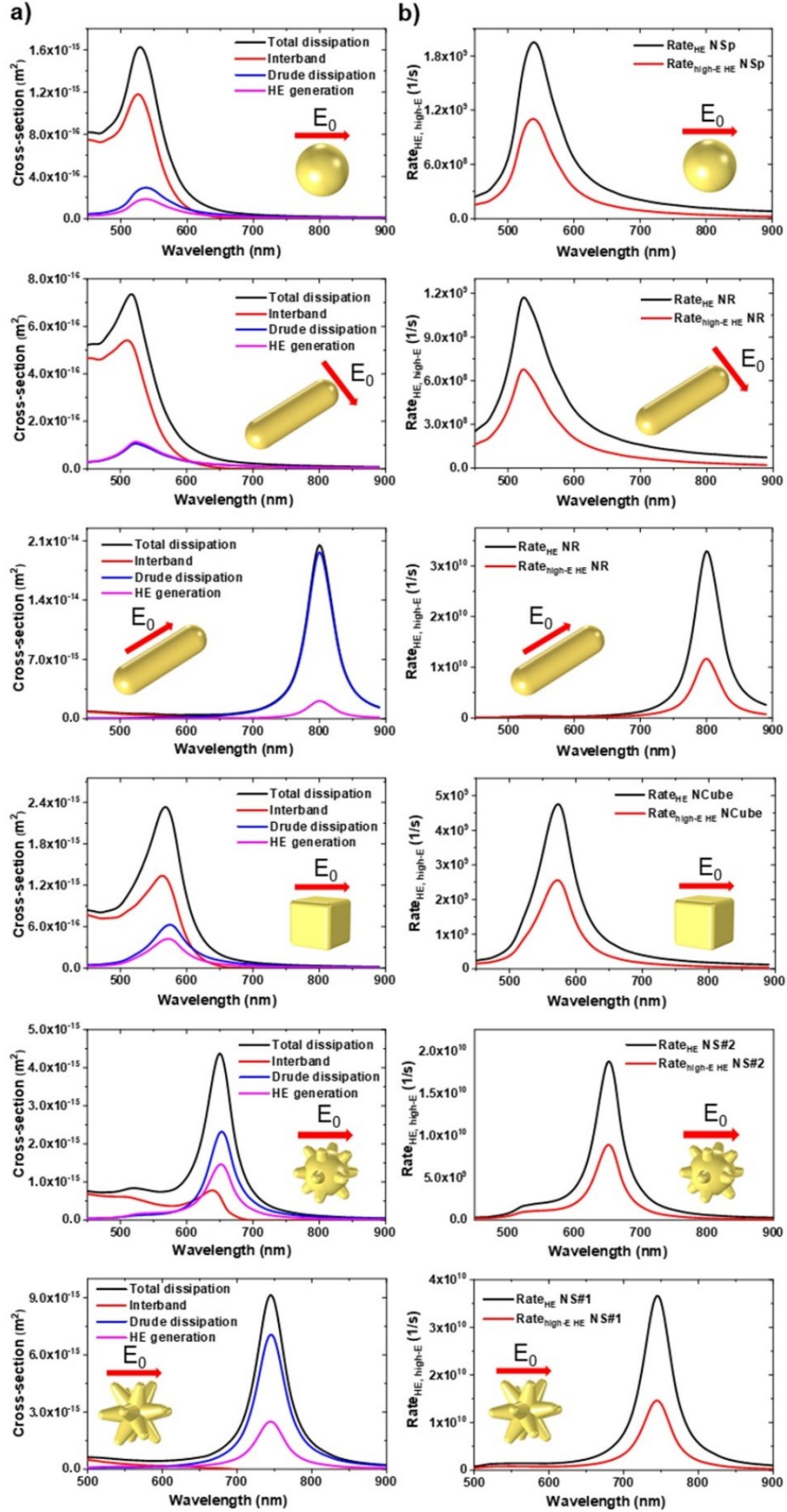

**Figure 3**. Optical and optoelectronic properties of Au NCs for a polarized excitation. a) Optical cross-sections of various gold NCs for different dissipation mechanisms in NCs. The light polarization is shown by the red arrow; $R_{\text{eff}} = 15\,\text{nm}$ for all NCs. b) Hot-electron generation

rates in the NCs. Here we show both the total rates and the rates for the over-barrier electrons; the energy barrier height is 1 eV.

**2.2 Hot-electron generation mechanisms.**

In bulk, predominantly, the low-energy carriers are excited because the excitation of the high-energy electrons is forbidden due to the conservation of the linear momentum (see problem 3 in Ch. 15 in ref. [43]). We regard such carriers as Drude electrons,[14,15,22] which are responsible for the coherent time-oscillating plasmonic currents leading to the plasmon resonance.

Of course, the main difference from the bulk case for the HE generation in NCs is the impact of the surface. Near the surface, the continuity of the lattice is broken, and correspondingly the conservation law of the linear momentum is not fulfilled, hence the hot carriers may be efficiently generated [44]. In the current literature, one can see numerous examples of the excitation of high-energy electrons in small metal nano-objects.[15,18,35,44,45] We note that the picture in which incident light at the initial absorption step generates mainly carriers with high energies (used for example in ref. [46]) is physically incorrect. One can find the correct picture and related explanation for the bulk case in a classical textbook by Ziman [47] (Ch. 8), and, for the discussion on NCs, one can see our papers of refs. [14,15,22]. The plasmonic wave is composed of coherent low-energy excitations (electron-hole pairs) near the Fermi surface. The acceleration creates those excitations in the Fermi sea, and the resulting plasmon frequency is a parameter incorporating only classical constants (see below for more discussion). The above picture of the dynamic response of conducting solids (the so-called Drude–Sommerfeld model) is a framework commonly presented in university courses. Kinetic theories supporting the Drude-electron picture of the plasmon oscillation can be found in refs. [2,47–49]; the above theories are based on various approaches, including classical, semiclassical, and quantum formalisms. For example, the classical Ziman's text (ref. [47], p. 279, Eq. 8.86) gives the following distribution of excited electrons in the plasmonic wave:

$$\delta f(\mathbf{p},t) = -\frac{\partial f^0(E)}{\partial E}\frac{e\tau \mathbf{v}\cdot \mathbf{E}_0}{1-i\omega\tau}e^{-i\omega t},$$

where $f^0(E)$ is the equilibrium Fermi distribution, $\tau$ the relaxation time, and $\mathbf{E}_0$ is the external electric field. The above equation tells us that the nonequilibrium electrons in a plasmonic wave in a 3D electron gas occupy a narrow energy interval near the fermi level. The width of this interval is $\sim 6k_BT$. This result is commonly found in textbooks on solid-state and condensed-matter physics, including the classical text by Ziman [47] (Ch.8), a recent excellent text by Tanner[2] (Ch.8), and a classical text by Kittel with exemplary quantum solid-state problems[50] (Ch.16).

One more note should be made regarding the 2nd-order photon-assisted processes that can create some number of energetic electrons in bulk. Phonon scattering is the leading mechanism at room temperature for the Drude electrons and governs the relaxation rate, $\gamma_D$, in the Drude dielectric constant.[51] However, those phonon processes are of 1st order in terms of the electron-phonon interaction. In that sense, the electron-phonon scattering is a crucial element of electronic dynamics. As it was explained above, Drude electrons, forming the coherent currents in the excited plasmon wave, carry low excitation energies, $\sim k_BT$. In addition to Drude electrons, there is a possibility to create carriers in a 3D crystal with high excitation energies, $\sim \hbar\omega$, via electron-phonon scattering; indeed, the phonon wave breaks the translation symmetry of the lattice, and the electron momentum conservation can be violated in this way. To describe this process, we need to apply 2nd-order perturbation theory, and, correspondingly, such phonon-assisted process should be considered a 2nd-order effect. It is known that these 2nd-order phonon-scattering processes in bulk are weak, and it is challenging to observe them experimentally.[8] Simultaneously, the HEs generated near the surface via the surface quantum mechanism and the interband excitations in a NC are observable and represent strong 1st-order physical effects.[44] Why are the above surface-HE and interband-HH processes so strong and measurable? (1) A NC may have strong plasmonic hot spots near the surface, which play the role in HE generation sites. (2) Surface-generated HEs can be extracted to an acceptor or transferred directly to the liquid through the surface. Surface charge acceptors, realized in recent experiments, are various, for example: $TiO_2$ clusters,[28] semiconductor films,[8] $PbO_2$ layers,[30] etc. Concerning the interband HHs, we already see that their contribution is very strong simply from looking at the bulk empirical dielectric constant; the physical reason for these prominent transitions is found directly in the periodicity of the crystal lattice. The optical absorption by the interband transitions dominates in the interval $\lambda < 600\,\text{nm}$. Our present theory does not directly incorporate the photon-induced mechanism of the HE generation in bulk, since this mechanism is weak and challenging to observe. We also note that this weak 2$^{\text{nd}}$-order channel of scattering is included in the empirical Drude relaxation constant, $\gamma_D$.

Now we look at the physical contributions to the absorbed power, $P_{\text{abs}}$. Considering the main microscopic mechanisms of dissipation in a plasmonic NC (Figure 2c,d), one should split the absorbed power into the following terms:[14,15,20]

$$P_{\text{abs}} = P_{\text{Drude}} + P_{\text{interband}} + P_s, \qquad (4)$$

where $P_{\text{Drude}}$, $P_s$, and $P_{\text{interband}}$ are the Drude, surface-scattering, and interband contributions, correspondingly. Each term in Equation (4) represents a specific pathway of energy dissipation in a metal NC (Figure 2c,d)

The first and second terms are the fractions of the energy dissipation caused by Drude and interband processes ($P_{\text{Drude}}$ and $P_{\text{interband}}$). Drude dissipation can be computed from the semiclassical Boltzmann theory, which is based on the free-electron model and the relaxation time approximation.[2,47] In a typical metal, the relaxation time at room temperature comes mostly from collisions of the conduction electrons with lattice phonons.[51] The second term ($P_{\text{interband}}$) in Equation (4) originates from the interband transitions in a noble metal; for gold, it becomes active for $\lambda < 680\,\mathrm{nm}$. Typically, the Drude and interband contributions (see Equation (4)) are the main terms responsible for absorption of optical energy in a plasmonic NC; however, in very small NCs, the surface scattering term should dominate.[15] Qualitatively, the origin of the Drude dissipation can be explained in the following semiclassical way. The external electric field provides extra energy to the mobile electrons via classical acceleration, and this extra energy is relatively small in the linear excitation regime. Then, the excited electrons can occupy the states only in the vicinity of the Fermi level; however, since those electrons receive extra energy, they transfer it to the lattice via phonon scattering. Before the collision event, a mobile electron (its center of mass) is moving in space according to classical Newton's law and its energy oscillates. The kinetic Boltzmann equation provides us with a robust quantitative description leading the dielectric function in the well-known Drude form:[2]

$$\varepsilon_{\mathrm{Drude}}(\omega) = 1 - \frac{\omega_{\mathrm{p}}^2}{\omega(\omega + i\gamma_{\mathrm{D}})},$$

where $\gamma_{\mathrm{D}}$ is the Drude broadening parameter and $\omega_{\mathrm{p}}$ is the plasmon frequency,

$$\omega_{\mathrm{p}} = \sqrt{\frac{e^2 n_0}{\varepsilon_0 m}}.$$

Here $n_0$, $m$ and $\varepsilon_0$ are the electron density, the electron's mass, and the dielectric permittivity of free space, respectively. First, we observe that all quantities in the equation for $\varepsilon_{\mathrm{Drude}}(\omega)$ are of a classical nature, and quantum mechanics seems to be irrelevant. However, the Drude function can also be derived using both quantum and semiclassical formalisms applied directly to the electronic Fermi sea, which is an essentially quantum concept.[2,47–49]

Figures 2c illustrate the bulk processes. The generation rates for the Drude and interband carriers should be computed using quantum kinetic theories.[14,15,22] The Drude electrons and holes (**DE** and **DH** in Figure 2e) are generated near the Fermi surface, and the corresponding spectral rate shows an intensive peak-dip structure near the Fermi level (see Figure 2e). The interband processes lead to the generation of high-energy d-holes and low-energy electrons in the sp-band (EEdB), as shown on the inset of Figure 2e. Energy efficiencies for the Drude and interband generations can be comparable depending on the shape of a NC.[14]

Finally, we look at the most interesting mechanism for us. All confined geometries possess a peculiar process - the surface scattering effect, leading to high-energy electrons generation.[44] These nonthermalized high-energy carriers are regarded here as hot electrons. The energy distribution of hot carriers is shown in Figure 2e. Intraband HEs and HHs occupy the wide interval of $E_F - \hbar\omega < E < E_F + \hbar\omega$ (inset of Figure 2e). Those carriers show a relatively small population in Figure 2e, and the corresponding energy efficiencies for the generation of these carriers are not appreciable.[14,15] The corresponding term in the total absorption rate (Equation (4)) is denoted as $P_s$. Those HEs and HHs are used for surface photoreactions [28].

In our previous publications, we developed a convenient computational formalism for obtaining the hot carrier rates.[14,15] Here we treat relatively large NCs, for which the size-quantization effects are not noticeable, and the local dielectric constant approach can be applied. First, we express the dielectric constant in the following way:

$$\varepsilon_{\text{metal}}(\omega) = 1 + \Delta\varepsilon_{\text{interband}}(\omega) - \frac{\omega_p^2}{\omega(\omega + i[\gamma_D + \gamma_s])},$$
$$\Delta\varepsilon_{\text{interband}}(\omega) = \varepsilon_{\text{metal, bulk}}(\omega) - 1 + \frac{\omega_p^2}{\omega(\omega + i\gamma_D)}, \qquad (5)$$

where $\varepsilon_{\text{metal, bulk}}(\omega)$ is the experimental dielectric constant for the bulk material.[52] Furthermore, $\gamma_s$ is the effective plasmonic broadening due to the electron scattering mediated by the surface; this broadening is a quantum parameter. The final equations to compute have the forms:

$$P_{\text{interband}} = \text{Im}\,\Delta\varepsilon_{\text{interband}}(\omega)\cdot\varepsilon_0\frac{\omega}{2}\int_{NC} dV\,\tilde{\mathbf{E}}_\omega\cdot\tilde{\mathbf{E}}_\omega^*,$$

$$P_{\text{Drude}} = \frac{\omega_p^2}{\omega^3}(\gamma_D)\cdot\varepsilon_0\frac{\omega}{2}\int_{NC} dV\,\tilde{\mathbf{E}}_\omega\cdot\tilde{\mathbf{E}}_\omega^*, \qquad (6)$$

$$P_s = \frac{\omega_p^2}{\omega^3}(\gamma_s)\cdot\varepsilon_0\frac{\omega}{2}\int_{NC} dV\,\tilde{\mathbf{E}}_\omega\cdot\tilde{\mathbf{E}}_\omega^*.$$

We note that the equations (Equation (6)) were derived under the assumption $\omega^2 \gg (\gamma_D+\gamma_s)^2$, which applies well to the typical plasmonic NCs.[14] The quantum broadening $\gamma_s$ is a size- and shape-dependent parameter, and it should be found computationally from a nonlinear integral equation, self-consistently.[14] We provide details in the Supporting Information. Correspondingly, the optical cross-sections for the above processes read as

$$\sigma_{\text{Drude}} = \frac{P_{\text{Drude}}}{I_0}, \qquad \sigma_s = \frac{P_s}{I_0}, \qquad \sigma_{\text{interband}} = \frac{P_{\text{interband}}}{I_0}.$$

For the solution case, our NCs have random orientations. Therefore, we need to average the cross-section in the following fashion:

$$\sigma_{\text{av}} = \frac{\sigma_{(k_x|E_y)}+\sigma_{(k_x|E_z)}+\sigma_{(k_y|E_x)}+\sigma_{(k_y|E_z)}+\sigma_{(k_z|E_x)}+\sigma_{(k_z|E_y)}}{6}, \qquad (7)$$

where $k_x$, $k_y$, and $k_z$ are the wavevectors oriented along the three orthogonal axes, and $E_x$, $E_y$, and $E_z$ are the corresponding electric-field polarizations of incident light.

### 2.3 Hot-electron maps.

The integrated and local HE generation rates for the three dissipation processes were introduced by us in ref. [14]. Do the HE generation rates carry additional information as compared to absorption? The answer is yes, they do. (1) The absorption spectrum is an integrated, nonlocal property, while the HE generation maps are position-dependent and provide us with an idea of where and how hot carriers are generated. (2) Furthermore, the absorption spectrum is calculated with classical electrodynamics, whereas the HE rates explicitly include $\hbar$ and $E_F$. (3) While absorption includes dissipation mechanisms altogether, the HE rates give individual contributions. (4) Finally, the HE rates include additional quantum factors coming from the transition matrix elements of the form of $(\hbar\omega)^{-3}$. We found earlier that these factors play a key role in enhancing HE generation in the near-IR interval when the quantum $(\hbar\omega)^{-3}$ factor rapidly grows. [8,11,53]

The HE generation rate ($Rate_{HE}$) defines the total number of HEs excited inside a NC per second. The second parameter ($Rate_{high-E\ HE}$) is defined for the high-energy HE; it gives the rate of excited electrons having over-barrier energies (Figure 2d).[14,54]

$$Rate_{HE} = \frac{1}{4} \times \frac{2}{\pi^2} \times \frac{e^2 E_F^2}{\hbar} \frac{1}{(\hbar\omega)^3} \int_{S_{NC}} |E_{\omega,\ normal}(\theta,\varphi)|^2 ds,$$

$$Rate_{high-E\ HE} = \frac{1}{4} \times \frac{2}{\pi^2} \times \frac{e^2 E_F^2}{\hbar} \frac{(\hbar\omega - \Delta E_{bar})}{(\hbar\omega)^4} \int_{S_{NC}} |E_{\omega,\ normal}(\theta,\varphi)|^2 ds,$$

(8)

where $E_{\omega,\ normal}(\theta,\varphi)$ is the normal component of the electric field on the surface, inside a NC; $E_F = 5$ eV is the Fermi energy for gold. The integral is taken over the surface; $\Delta E_{bar}$ is the injection barrier energy; in our computations, we will use $\Delta E_{bar} = 1$ eV, which is a typical number for the Au-TiO$_2$ Schottky-barrier. As one can see from Equation (8), $Rate_{high-E\ HE}$ describes the generation of electrons with high (over-barrier) energies, i.e., with $E > \Delta E_{bar}$. In a solution setting, the rates (Equation (8)) should be averaged over twelve light configurations.

One of the central goals of this paper is to look at the local rates of HE generation at the surface: [14]

$$r_{\text{HE, surface}}(\mathbf{r}) = \frac{d^2 N_{\text{HE}}}{dtds} = \frac{1}{4} \times \frac{2}{\pi^2} \times \frac{e^2 E_F^2}{\hbar} \frac{1}{(\hbar\omega)^3} \left|E_{\omega,\text{normal}}(\theta,\varphi)\right|^2,$$

$$r_{\text{high-E HE, surface}}(\mathbf{r}) = \frac{d^2 N_{\text{high-E HE}}}{dtds} = \frac{1}{4} \times \frac{2}{\pi^2} \times \frac{e^2 E_F^2}{\hbar} \frac{(\hbar\omega - \Delta E_{\text{bar}})}{(\hbar\omega)^4} \left|E_{\omega,\text{normal}}(\theta,\varphi)\right|^2. \tag{9}$$

As one can see, the units of the above rates are $1/(s \cdot m^2)$. Correspondingly, the d-hole generation process, which is a bulk mechanism, is given the rate related to the unit of volume:

$$r_{\text{d-holes, bulk}}(\mathbf{r}) = \frac{d^2 N_{\text{d-holes}}}{dtdV} = \frac{1}{\hbar\omega} \text{Im}[\Delta\varepsilon_{\text{interband}}]\varepsilon_0 \frac{\omega}{2} \mathbf{E}_\omega \cdot \mathbf{E}_\omega^* . \tag{10}$$

The local power used for the d-hole generation should be written as:

$$P_{\text{d-holes, bulk}}(\mathbf{r}) = \frac{d^2 P_{\text{d-holes}}}{dtdV} = \text{Im}[\Delta\varepsilon_{\text{interband}}]\varepsilon_0 \frac{\omega}{2} \mathbf{E}_\omega \cdot \mathbf{E}_\omega^* .$$

Finally, we look at the local power map for the Drude dissipation, i.e., for the local dissipation power related to the plasmonic currents:

$$P_{\text{Drude, bulk}}(\mathbf{r}) = \frac{d^2 P_{\text{Drude, bulk}}}{dtdV} = \frac{\omega_p^2 \cdot \gamma_D}{\omega^3} \varepsilon_0 \frac{\omega}{2} \mathbf{E}_\omega \cdot \mathbf{E}_\omega^* . \tag{11}$$

Correspondingly, the total d-hole rates should be given by the volume integrals:

$$\text{Rate}_{\text{d-holes, bulk}} = \varepsilon_0 \frac{1}{\hbar\omega} \text{Im}\,\Delta\varepsilon_{\text{interband}}(\omega) \frac{\omega}{2} \int_{NC} dV\, \mathbf{E}_\omega \cdot \mathbf{E}_\omega^*,$$

$$P_{\text{d-holes, bulk}} = \text{Im}(\varepsilon_{\text{interband}})\varepsilon_0 \frac{\omega}{2} \int_{NC} dV\, \mathbf{E}_\omega \cdot \mathbf{E}_\omega^* . \tag{12}$$

Supporting Information contains all physical parameters entering the above equations. Below we will use Equation (8-12) to compute various generation mechanisms for a few prominent NC geometries.

Our approach is COMSOL-based and directly involves the band structure of the metal via the local dielectric constant $\varepsilon_{\text{metal}}(\omega)$. This allows us to compute NCs with complex shapes and large sizes. The band structure of gold is reflected in our theory through the term $\Delta\varepsilon_{\text{interband}}(\omega)$, which has a threshold at $\lambda = 680$ nm. In other words, our approach is based on the empirical dielectric function that should take care of the band-structure effects. Some theoretical *ab initio* approaches, which are widely celebrated, are Density Functional Theory (DFT) and Time Dependent (TD)-DFT. In particular, the TD-DFT is a powerful technique directly solving the time-dependent Schrödinger equation incorporating the atomistic Kohn-Sham Hamiltonian.[55] One recent example of a TD-DFT study can be found in Ref. [56]; this paper treats a Ag cluster and explicitly computes the HE distribution, which reveals both the high-energy carriers and the low-energy peaks near the Fermi level (i.e., Drude electrons). In the Supporting information, we discuss some other methods to approach the HE problem such as second-quantization Hamiltonians for the plasmonic degrees of freedom.

To conclude this section, we should comment on the nonequilibrium electron populations in an optically driven NC. For the integrated generation rates (Equation (8,12)) the self-consistent relaxation theory was recently developed by us in *r*ef.[14] The HE rates computed here play the role *in* local sources of HEs. Furthermore, the high-energy electrons and holes populations also depend on the e-e relaxation time, $\tau_{e-e}(E)$;[14] this strongly energy-dependent parameter obeys in 3D systems the famous Landau's law: $\tau_{e-e}(E) \sim (E-E_F)^{-2}$. The present study does not include the relaxation processes, but we rather look at the other critical elements of the HE generation effect – the local surface and bulk generation rates. In other words, we show where and how (and via what quantum processes) the optical field generates energetic carriers, which can be used for photochemistry, in a photodetector, or for other applications. Another technologically important relaxation channel is cha*r*ge transfer through the surface to semiconductor accept*o*rs or molecules in *a* liquid. Our formalism assumes that *the* transfer of charge weakly *a*ffects the HE processes since our theory deals with the weak excitation limit, which is typical for the plasmonic photochemical studies in the literature. The number of extracted electrons is much smaller that the equilibrium number of electrons in a NC, and, therefore, we can neglect the cha*r*ge-transfer effect for such a linear regime. Some elements of the cha*r*ge transfer theory can be found in our previous studies of *r*efs. [14,57]. Certainly, the HE transport and relaxation processes in NCs with complex shapes need further development in terms of theory.

The generation rates computed here can be applied directly to certain photochemical situations. One example is the chiral growth in ref. [30] that reflects a surface pattern of the electromagnetic field; our recent theory[31] reproduces well such experimental growth (Figure 1b,c). Another example is an experiment with a comparative study of Au NSps, NRs, and NSs; our HE theory based on the generation rates shows that the NS geometry is the most efficient, due to the hot spots on the NS tips. Finally, our local formalism and generation maps explain the experimental results of ref. [28] and should guide more experiments (like a recent study of ref. [32]). For the situations where the transport could essentially influence the photochemical picture, our theory should be able to predict several experimental properties of immediate importance:

1. The anisotropy and uniformity of the photoreactions on the surface and in solution.
2. The NC geometries that are efficient in terms of the hot electron generation rates.
3. The spectral position-dependent and integral rates of the main hot-carrier mechanisms (such as the intraband HE generation and the interband d-hole excitation) in different complex geometries.

**2.4 Temperature formalism.**

For certain chemical reactions, an increase of chemical rates can also appear due to photoheating. Now we write down the related local function $\Delta T(\mathbf{r},t) = T(\mathbf{r},t) - T_0$, where $\Delta T$ is the local temperature increase, and $T_0$ is the ambient one, recorded at large distances. In the next step, we solve this problem using the heat diffusion equation [58]:

$$\rho(\mathbf{r}) \cdot c(\mathbf{r}) \frac{\partial T(\mathbf{r},t)}{\partial t} = \vec{\nabla} \cdot (k(\mathbf{r})\vec{\nabla} T(\mathbf{r},t)) + q_{local}(\mathbf{r},t) \qquad (13)$$

where $T(\mathbf{r},t)$ is the temperature as a function of the coordinate $\mathbf{r}$ and the time $t$. $\rho(\mathbf{r})$ is the mass density, $c(\mathbf{r})$ is the specific heat capacity, and $k(\mathbf{r})$ is the thermal conductivity. $q_{local}(\mathbf{r},t)$ denotes the local heat generation due the dissipation of light in the metal NC.[59]

$$q_{local}(\mathbf{r},t) = \varepsilon_0 \frac{\omega}{2} |\mathbf{E}_\omega(\mathbf{r})| \cdot \text{Im}(\varepsilon_{metal}) \cdot F_{pulse}(t)$$

Here $\mathbf{E}_\omega(\mathbf{r})$ is the electric field inside the NPs, which we have already defined in Equation (1). In the CW illumination regime, $F_{pulse}(t) \equiv 1$. We choose water as a surrounding environment, where the NCs are submerged. One may find the material coefficients used for the optical and thermal computations in Tables S1,S2 (Supporting Information).

In our numerical calculations (see the following figures), we found that $T(\mathbf{r})$ is nearly uniform inside the NCs.[60–62] This is due to the very high thermal conductivity of the metal: $k_{Au} \gg k_{water}$. More precisely, the parameter that should control the uniformity of the local temperature inside the NC, is $\tau_{T,Au} / \tau_{transfer}$, $\tau_{T,Au}$ and $\tau_{transfer}$ are the characteristic time of thermal diffusion inside the Au NC and the time of heat release from the NC to the matrix, respectively. In the Supporting Information, we estimate these time parameters and observe that $\tau_{T,Au} / \tau_{transfer} \sim 0.006$ for a NC of $R_{eff} = 15 \text{ nm}$. The above property, i.e., $\tau_{T,Au} / \tau_{transfer} \ll 1$, ensures that the phototemperature should be nearly uniform within the metal component.

However, the above property (i.e., $T(\mathbf{r}) \approx const$ within the NC) has certain limitations, of course (Supporting Information). Briefly, this property is found in relatively small NCs for which the near-field approximation is valid. For large NCs and metastructures, the temperature gradients within the metal can be strong and drive surface photochemistry locally.[63,64] One insightful case is a "*shuriken*" metastructure with long arms where the phototemperature is strongly nonuniform and the photochemical processes are position dependent.[64]

We also should consider our results in the context of the lattice and electronic temperatures ($T_L$ and $T_e$).[65,66] Such a two-temperature (2T) model was widely employed to describe time-dependent phenomena in NCs.[10,22] Our formalism based on the local differential equation (Equation (13)) has a single temperature that should be considered as the common temperature of the lattice and electrons, since under CW excitation $T_L \approx T_e$. This condition comes from the fact that the electron-phonon relaxation in a metal is fast. Namely, $\tau_{e-ph} \ll \tau_{transfer}$, where $\tau_{e-ph}$ the electron-phonon relaxation time in a NC. For Au NCs, experiments typically show: $\tau_{e-ph} \sim 0.5\,\text{ps}$.[7,12] Considering our NCs with $R_{eff} = 15$ nm, we have for the transfer time: $\tau_{transfer} \sim 310\,\text{ps}$. Then, we find that $\tau_{e-ph} / \tau_{transfer} \sim 1.6 \times 10^{-3} \ll 1$. Therefore, in this case the electronic and lattice temperatures should be very close to each other, and the single temperature approach is fully applicable. In other words, $T = T_L \approx T_e$. As the last comment on our thermal treatment, we should mention that the calculated parameter $\Delta T(\mathbf{r})$ is the local phototemperature created by a single NC. In a typical experiment with NC solutions, the actual temperature around a single NC is made of two contributions: $\Delta T_{tot}(\mathbf{r}) = \Delta T(\mathbf{r}) + T_{coll}$, where $T_{coll}$ is the collective excess temperature created by the rest of the NC ensemble in a solution;[60,61,67] this term does not vary noticeably in space in the vicinity of a given NC. And, typically, $T_{coll} \gg \Delta T(\mathbf{r})$. If a plasmonic photoreaction at the NC surface is thermally driven, the total rate of the reaction can be controlled by $T_{coll}$. Simultaneously, the anisotropy of the reaction on the surface is controlled by $\Delta T(\mathbf{r})$.

## 3. Results for the directional excitation: Substrate setting.

Now we look at the computational data obtained with the help of COMSOL Multiphysics. We assume that our NCs are on a substrate and illuminated from one direction. To compare different shapes consistently, all NCs in our calculations will have the same volume, characterized by the effective radius.

$$V_{NC} = \frac{4\pi}{3} R_{eff}^3$$

In all subsequent figures, we present NCs with $R_{eff} = 15$ nm. Assuming a glass substrate and water as two adjacent media, we adopt a model with a uniform matrix of averaged dielectric constant ($\varepsilon_{env} = 2$). Indeed, the optical constant of glass and water are close, and this approach should provide us with reliable results.[68]

Figure 3a,b show both optical cross-sections and HE generation rates. The optical absorptions contain the contributions from the interband transitions, Drude-like dissipation, and surface-mediated electron scattering (HE generation). As expected, the plasmon resonance shows excellent tunability with the NC shape. We also observe that the NR geometry is strongly anisotropic, whereas the NSps, NCubes, and NSs show nearly isotropic cross-sections and integrated rates. We also note that the d-hole interband process is active mostly for the NSp geometry, appearing in the interval of $\lambda < 680$ nm.

Next, we compute the microscopic photophysical properties under polarized excitation (Figures 4 and 5). First, we see a striking difference between the surface maps for the phototemperature and the HE generation in Figure 4. The surface phototemperature is nearly uniform over the NC, whereas the HE rates show very strong variance and anisotropy. The reason comes directly from the physical mechanisms underlying them. The steady state phototemperature originates from the heat diffusion process; therefore, since the thermal conductivity is really high in the metal, this magnitude is nearly uniform within the crystal volume. More precisely, here we should compare two thermal diffusion times, for which we see that $\tau_{T,Au}/\tau_{transfer} \ll 1$, i.e., the thermal diffusion time inside the NC is much shorter than the time of heat escape from the NC (see the Supporting Information for estimates). Simultaneously, the local surface HE generation rate in Figure 4 is strongly anisotropic since it is given by the function $|E_{\omega,normal}(\theta,\varphi)|^2$ (Equation (9)). For the simplest NSp geometry, it reflects the polarization of the incident field (Figure 4a). For the other geometries in Figure 4, as expected, one can see the formation of plasmonic hot spots near the surfaces with strongly enhanced HE rates; the strongest hot spots occur for the NS case.[14,40] Since the distributions for the phototemperature and HE rate are so different, we should expect that the distributions of the local photochemical reactivity should also be very different for those mechanisms. Indeed, the experiments on surface photo-growth on a substrate show a striking variety of regimes. The photo-growth on a substrate in ref. [37] was interpreted as a purely plasmonic thermal mechanism. Indeed, the growth was isotropic,[37] and its speed correlated with the total absorption by the nano-antenna composed of three nanoparticles. Within a single element of the nanoantenna in ref. [37], the temperature distribution is expected to be uniform; however, the plasmonic hot-spot effect can strongly enhance the phototemperature on one of the elements. We note that similar conclusions for the phototemperature generation were made theoretically by us in Ref. [59]. A remarkably different photo-growth regime can be found in ref. [30]. This case is shown in Figure 1b; circularly polarized light under unidirectional excitation creates a prominent chiral distortion of the NC shape via surface photo-growth induced by plasmonic HEs. Such chiral growth appears in the above-mentioned paper [30] as an interplay of the chiral light and the anisotropic NC geometry. Regarding chiral growth, there are requirements for the symmetry of a NC: (1) if a NC is achiral, its geometrical cross section has to break rotational symmetry; i.e., a chiral distortion will not appear for spherical NCs;[31,54] (2) if a NC is chiral from the beginning, chiral photochemical effects (growth or surface chemistry) will be strongly enhanced.[32,54]

Nowadays, chiral plasmonic photochemistry is a very active area, rapidly expanding as one can see from refs [30,32,69,70].

Figure 5 gives a data summary for the mechanism involving the excitation of high-energy d-holes, in which incident photons promote the transitions from the d-band to the sp-band in gold (Figure 2c,d). The generation in this case takes place in bulk and, correspondingly, we plot cross sections of the 3D maps of the process. The local generation rate (power) is proportional to

$$\text{Im}(\Delta\varepsilon_{\text{interband}}) \cdot \mathbf{E}_\omega \cdot \mathbf{E}_\omega^*. \qquad (14)$$

The $\varepsilon_{\text{interband}}(\omega)$ function should be carefully extracted from the empirical dielectric constant by separating from the Drude contribution that dominates in the long wavelength limit (see Table S1 in the Supporting Information here and the formalism in ref. [14]). Even though the process is bulk-like, the generation rate can vary strongly inside a NC (Figure 5). Again, its shape matters: a spherical shape has a uniform rate distribution, whereas anisotropic NCs with hot spots show very strong variations of the local generation rates. An experimental study in ref. [17] (Figure 1e) presents an elegant method to separate the photoheating and the d-hole effects; in which the authors combined single-particle optical spectroscopy at different illumination wavelengths, transmission electron microscopy, and thermal characterization. It was found that the process of growth is isotropic and induced by the d-hole generation; the photothermal mechanism was excluded. This observation is fully consistent with our results in Figure 5 for the simplest NSp geometry; the pattern of the d-hole excitation is nearly uniform everywhere inside the NC. However, the above behavior of the NSps is drastically different from NCs with complex shapes (Figure 5). The local d-hole generation in anisotropic NCs in Figure 5 is strongly nonuniform. To illustrate this effect, we plot the d-hole rate along selected lines, which are located at different distances to the surface (0, 2.5, 5 nm). In a photochemical experiment, incident light first generates the d-holes in the bulk of the NC. Afterwards, these generated carriers with high energy propagate in all directions, and we are interested in the carriers reaching the surface. Due to their fast relaxation, a d-hole should lose its energy as it propagates away from the generation point, and this process is typically an exponential function of the travelled distance. Therefore, d-holes created near the surface have higher chances to reach the surface and to participate in surface photochemistry. Consequently, the regions with high rates near the surface (see the bulk maps in Figure 5) should indicate the corresponding increased reactivity at the surface.

Looking at the different NC shapes in Figure 5, we conclude that the highest numbers of excited d-holes are expected near the tips for NRs and NSs (the hot spot regions) and at the two opposite sides for the NCubes. Having strong anisotropy of the d-hole generation in the vicinity of the surface, one should expect a non-uniform distribution of the related reactivity, either driving NC growth or another photoreaction.

To conclude this section, we note that photochemistry on a substrate with unidirectional illumination shows some striking differences to that in a solution. However, these two settings also have much in common. For example, in both cases plasmonic hot spots play an important role;[28,71] Figures 1a,b,h present a few interesting examples of plasmonic hot-spot photoreactions. The advantage of the substrate setting lies in the possibility of studying single-particle spectroscopy and microscopy, and the latter one allows us to directly observe shape and size transformations [30,36,37] with the aid of various local microscopy means (TEM, SEM, AFM, etc.);

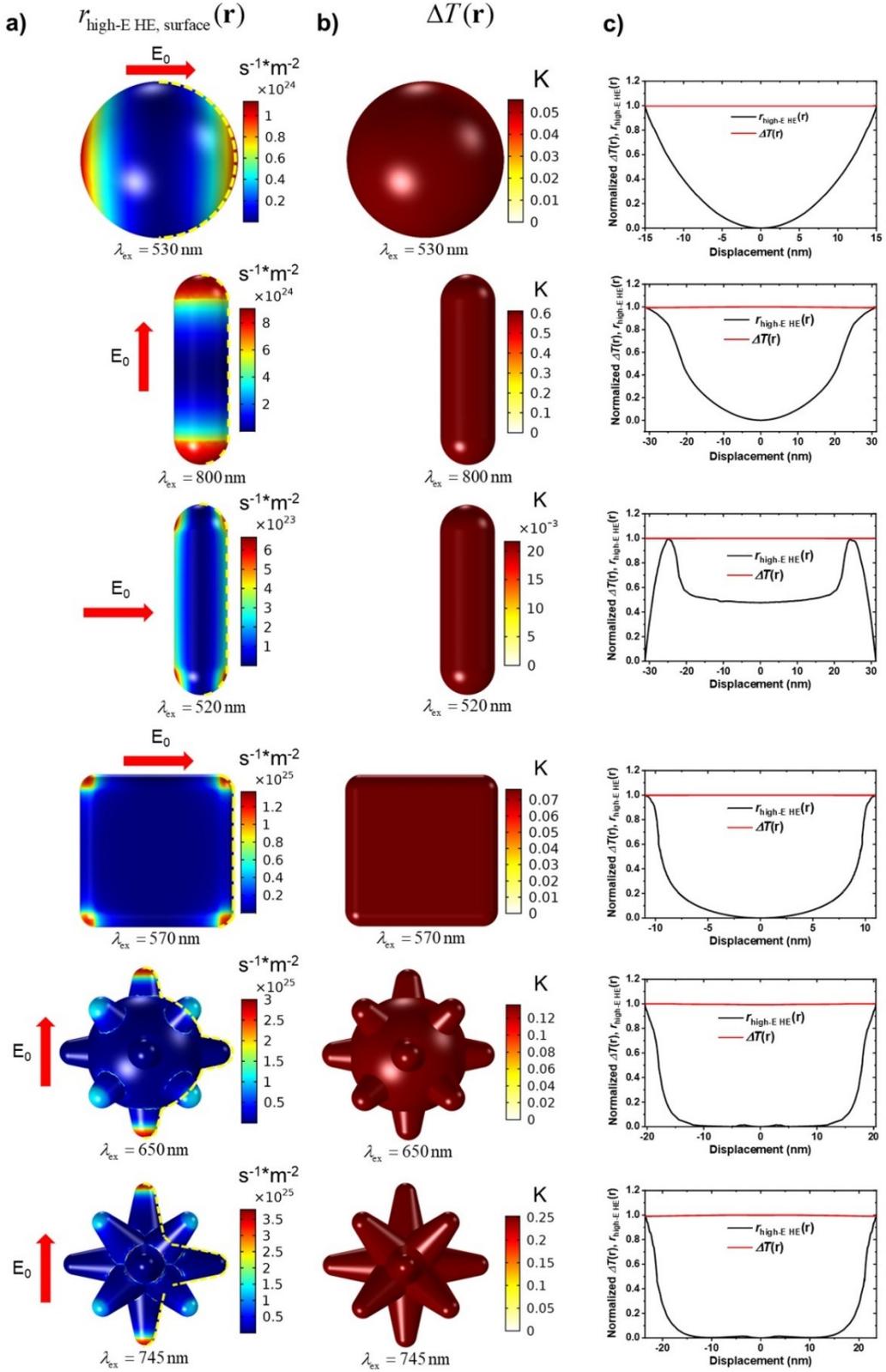

**Figure 4.** Computational data for the case of polarized excitation. a) Local distributions of the high-energy (over-barrier) HE generation rate; b) surface maps for the phototemperature. The excitation wavelength for each NC is taken at the plasmon peak. The polarization is shown by the red arrow. All NCs have an effective radius of $R_{eff} = 15\,\text{nm}$. c) These curves describe the anisotropy of the distributions of the temperature and the high-energy (over-barrier) HEs excitation. In panels (a), the yellow dashed lines show the spatial positions for the plotted rates and phototemperature. In (c), the normalization is done by dividing by the maximum value. The excitation wavelengths ($\lambda_{ex}$) are taken at the plasmon peaks for all shapes.

## Substrate setting with polarized excitation: The bulk maps of d-hole generation

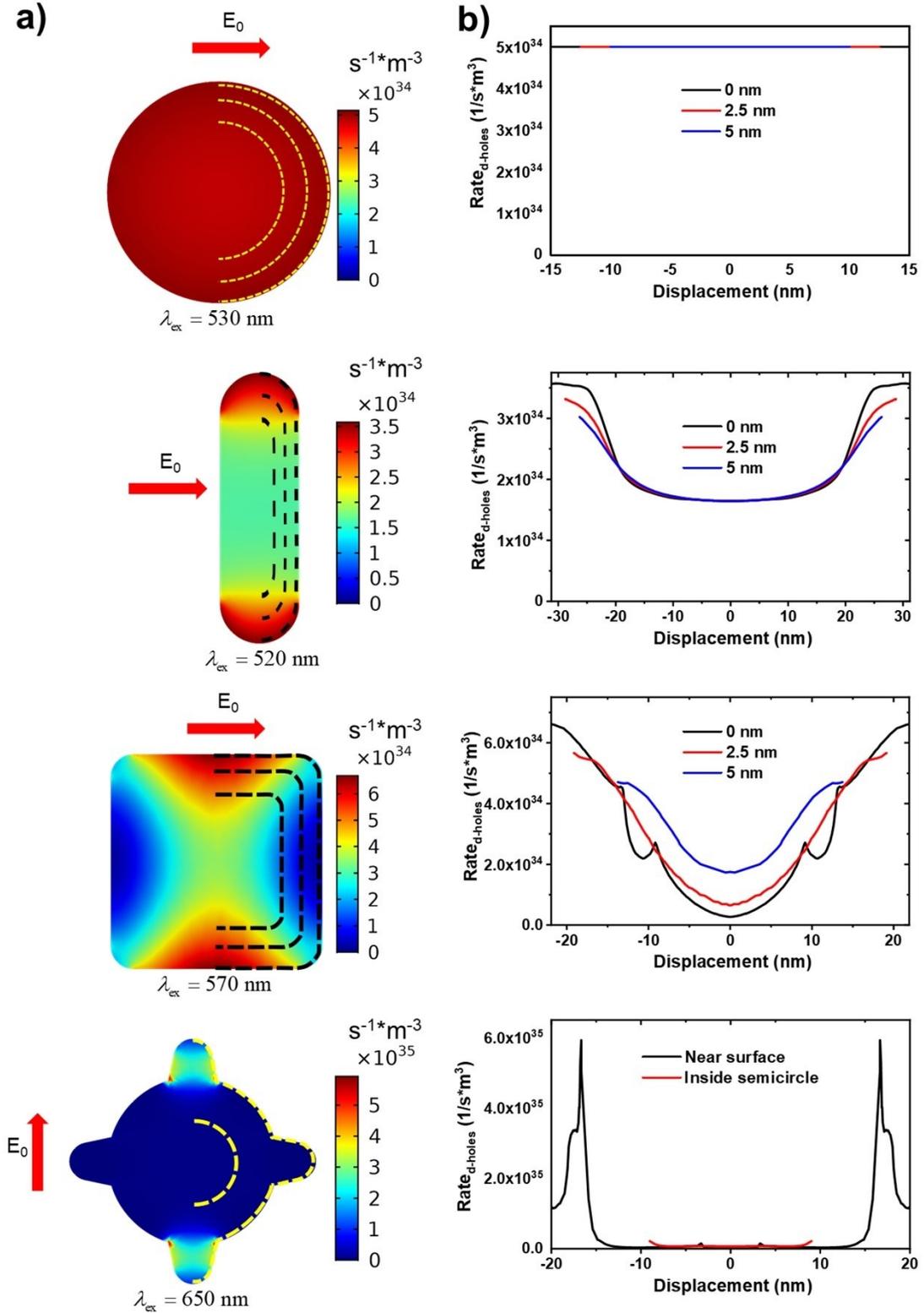

**Figure 5.** a) Local bulk maps of the d-hole generation rate in the NCs, under linearly polarized excitation. The polarization is shown by the red arrow. For the effective radius, we use

$R_{eff} = 15$ nm for all NCs. b) The d-hole generation rate near the surface and inside, plotted along the dashed lines noted in panels (a); the distances between the lines and the surface are 0, 2.5, and 5 nm. The excitation wavelengths ($\lambda_{ex}$) are taken at the plasmon peaks for all shapes.

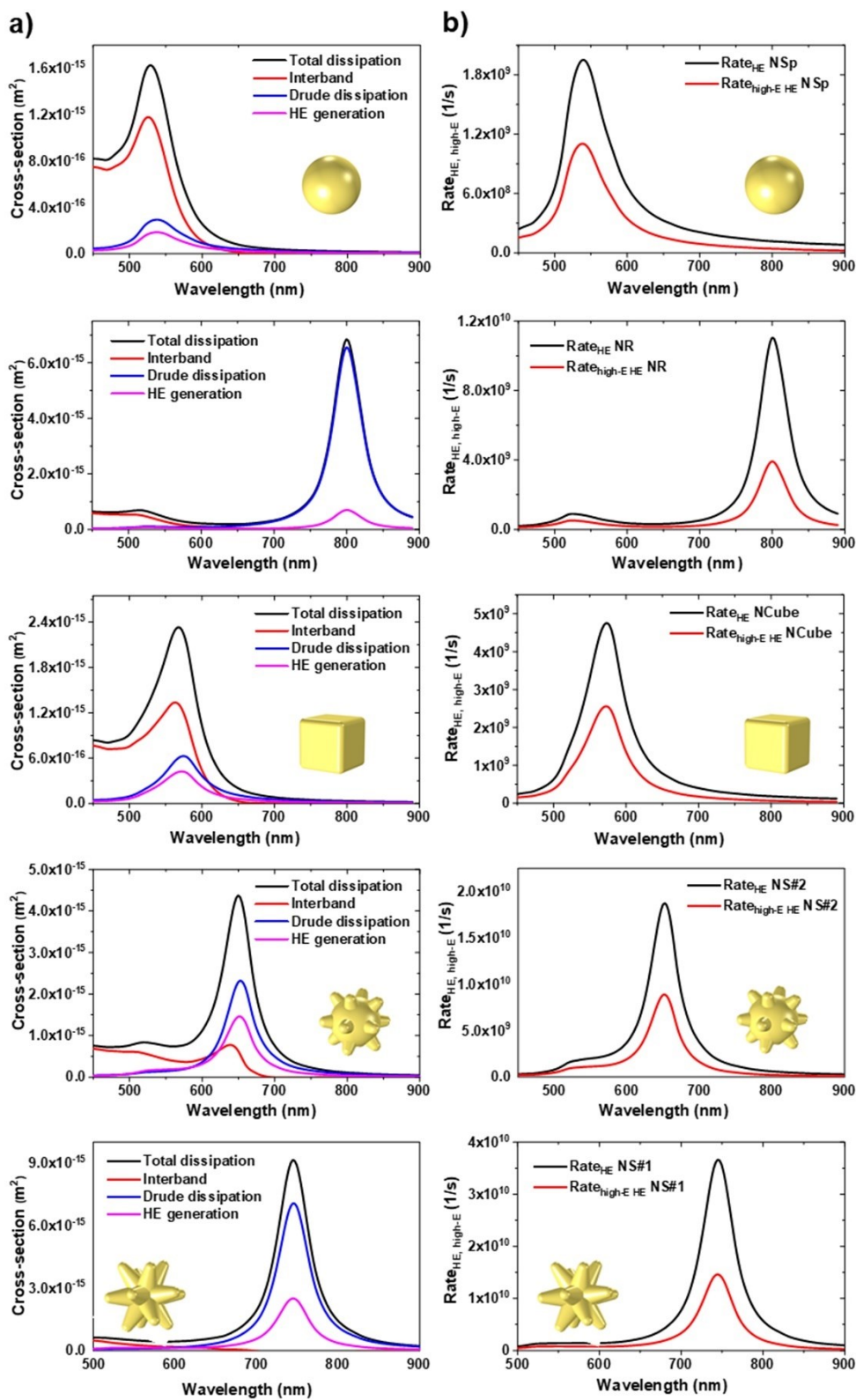

**Figure 6.** Optical and optoelectronic properties of Au NCs for non-polarized excitation (solution). The spectra are averaged over the different excitation configurations. a) Optical cross-sections of gold NCs for various dissipation mechanisms. The effective size for all NCs

is the same, $R_{eff} = 15\,nm$. b) Rates of HE generation. Hot-electron generation rates in the NCs. Here we show both the total rates and the rates for the over-barrier electrons; the barrier energy is 1 eV.

## 4. Hot-electron generation in a solution.

Experimentally, the most accessible and well-reported setting for plasmonic NCs is in solution, of course. Here, the two experimental settings, shown Figure 2a, represent two models, which include unidirectional excitation of a NC on a substrate (model 1) and randomly oriented NCs in a solution phase (model 2). Importantly, one can notice a few fundamental differences between these two models when considering surface photochemistry. Figures 4 and 5 show the substrate case, and Figures 6, 7, 8, and 9 present the data for the solution model. In the solution model, we illuminate our NCs from orthogonal directions and then find average values.

For the far-field cross-sections, it suffices to use the three directions (x, y, and z) and overall, six configurations, as reflected in Equation (7). For the near-field responses, such as the HE rates and phototemperature, one should use all six incidence directions and twelve configurations in total:[14]

$$\sigma_{av} = \langle \sigma(\mathbf{k},\mathbf{E}_0) \rangle_{6\,configurations}, \quad Rate_{av} = \langle Rate(\mathbf{k},\mathbf{E}_0) \rangle_{12\,configurations}.$$

Then, by using the averaged values, the whole picture of optoelectronic phenomena for NCs with various shape is recovered, including the energy efficiencies.[14]

$$Eff_{HEs} = \frac{P_s}{P_{abs} + P_{scat}} = \frac{Rate_{HEs}}{(P_{abs} + P_{scat})/\hbar\omega},$$

$$Eff_{High\text{-}E\,HEs} = \frac{P_s}{P_{abs} + P_{scat}} = \frac{Rate_{High\text{-}E\,HEs}}{(P_{abs} + P_{scat})/\hbar\omega},$$

$$Eff_{\text{d-holes}} = \frac{P_{\text{interband}}}{P_{\text{abs}} + P_{\text{scat}}} = \frac{Rate_{\text{d-holes}}}{(P_{\text{abs}} + P_{\text{scat}})/\hbar\omega}.$$

At this point, we can also define the photothermal efficiency, which involves the scattering power and describes the ratio of the thermal losses and total extinction:

$$Eff_T = \frac{P_{\text{abs}}}{P_{\text{abs}} + P_{\text{scat}}}. \quad (15)$$

Figures 6 and 7 show the results for the nonlocal properties, including optical, photoelectric, and photothermal parameters. In particular, Figure 7 provides us with the efficiencies of the shapes at the plasmonic peaks; a detailed discussion on the efficiencies can be found in our previous paper, ref. [14]. The efficiencies were computed using the nonlinear self-consistent approach based on the Kreibig's parameter, $\gamma_s$ (see Equation (5) and Ref. [14]). Below we list the characteristic physical properties of the kinetic mechanisms computed by us. (1) The interband HE generation originates from the inelastic surface-assisted scattering (the Kreibig's mechanism [44]). This process is strongly enhanced in the red and infrared spectral intervals, [8,11,22] and it is also strongly promoted at plasmonic hot spots.[39,40] In the red spectral interval, the plasmonic field-enhancement effect is especially strong because the plasmon resonance does not overlap with the interband transitions.[7,20] Another reason for the enhanced intraband HE generation in the red is the $(\hbar\omega)^{-3}$ factor in Equation (8) for the rate. The above factor is of quantum origin and arises from the matrix elements, which describe the breaking of linear moment conservation at the surface.[15] Such an optoelectronic response is widely used for nanostructured photodetectors [72]. (2) The d-hole processes are active for $\lambda < 680$ nm and dominate the absorption cross-section for the NSp at the plasmonic peak. (3) Drude carriers are nonequilibrium electrons with low excitation energies, which form the coherent plasmonic currents in a NC; those carriers are well described by the semi-classical Drude model. (4) The photothermal efficiencies ($Eff_T$, Equation (15)) are always high for small sizes, as also shown in Figure 7b; the reasons for such behavior are the fast relaxation of high-energy carriers and the low light-scattering cross-sections for small NCs.

Looking at the above hot-carrier mechanism (1-4), the intraband HE generation, i.e., Kreibig's surface mechanism,[44] is probably the most special process occurring in plasmonic nanostructures, and more so in complex-shape NCs with hot spots. Indeed, hot-spot driven photoreactions and anisotropic growth are prominent phenomena observed in several impressive experiments, which also utilized local microscopies (such as TEM, SEM, etc.).[19,49,73] Figure 1h,j shows two examples of such experimental work.

Finally, Figures 8 and 9 present the key results for the model of NCs in solution. First, we should look at the spherical NC, and we see the fundamental difference between the solution and substrate models (Figure 4a,c). Of course, the case with randomly oriented NSp has no surface and bulk anisotropies for the surface HEs and d-holes, whereas the substrate case in Figure 4a,c reflects the incident-field orientation. Importantly, the anisotropy of the HE generation remains very strong for all NCs with complex shapes, even in solution; as seen in Figures 8 and 9, the surface and bulk maps for the HE generation rates reflect plasmonic hot spots appearing at tips and apexes. However, the symmetry of the maps is higher for the solution case and reflects the NC topology. Whereas, for the polarized-excitation case, the symmetry of the spatial maps is lower, because of the incident field polarization (Figures 4 and 5). The above properties hold for all high-energy HE generation mechanisms, including surface HEs and d-holes. For the generation of low-energy Drude electrons, we also see a similar behavior - see the Supporting Information (Figures S2 and S3). Again, the temperature generation maps are nearly uniform for the solution case (Figure 8).

The spatial distributions of hot carrier excitation and phototemperature determine how the chemical reactivity is distributed over the surface of the NC and how it would grow in the right environment. Our photophysical formalism allows us to compute such properties. One example of photochemistry in solution is a recent work on plasmonic photometry with TiN-Pt hybrids;[38] the above study deals with the NCube geometry and demonstrates that both photochemical mechanisms (intraband HEs and photoheating) are active simultaneously. The NCube geometry in ref. [38] indeed has the hot spot areas located at the apexes of the cube (Figure 8a), and it is where most of the photochemical activity should take place. The role of hot spots for NC photocatalysis in a solution can be revealed by comparing different geometries and materials. Such a study was done experimentally in ref. [28,42] and theoretically in refs. [14,40]. To see the shape effects, the authors of ref. [28] compared the photocatalytic efficiencies of NSps, NRs, and NSs. The photocatalytic material efficiency, introduced by us in Refs. [14,28], was defined as $Rate_{chem} / m_{NC}$, where $Rate_{chem}$ and $m_{NC}$ are the photochemical rate and NC mass, respectively. Importantly, the best performance was found for the plasmonic NSs, because of the strong field-enhancement effect at the nanostar tips.[14,40,71]

Plasmonic d-hole photocatalysis is another ongoing research direction.[18,33,74,75] It has been studied mostly by using spherical NCs made of Au and Ag. Importantly, this kind of photochemistry of NCs does not directly follow the plasmonic-field enhancement in a NC, because the d-hole rate includes another important multiplier: $\text{Im}[\varepsilon_{interband}]$ (see Equation (12, 14)). The resulting curve for the rate receives a prominent shoulder in the blue region (see Figure 7a for the NSp case). This behavior agrees well with the experimental photocatalytic data reported in ref. [18]; we note here that our paper of ref. [14] has a detailed comparison between the experimental d-hole data,[18] d-hole theory, and field-enhancement factor. Remarkably, the d-hole photochemical processes in Au and Ag NSps allow us to realize a family of carbon-based photoreactions observed with the aid of single-particle SERS.[75]

Photochemical kinetic behaviors are always intricate and intriguing - one example is the recent observation of nonlinear and stochastic dynamics in plasmonic photocatalysis.[74]

## Summary of the shapes

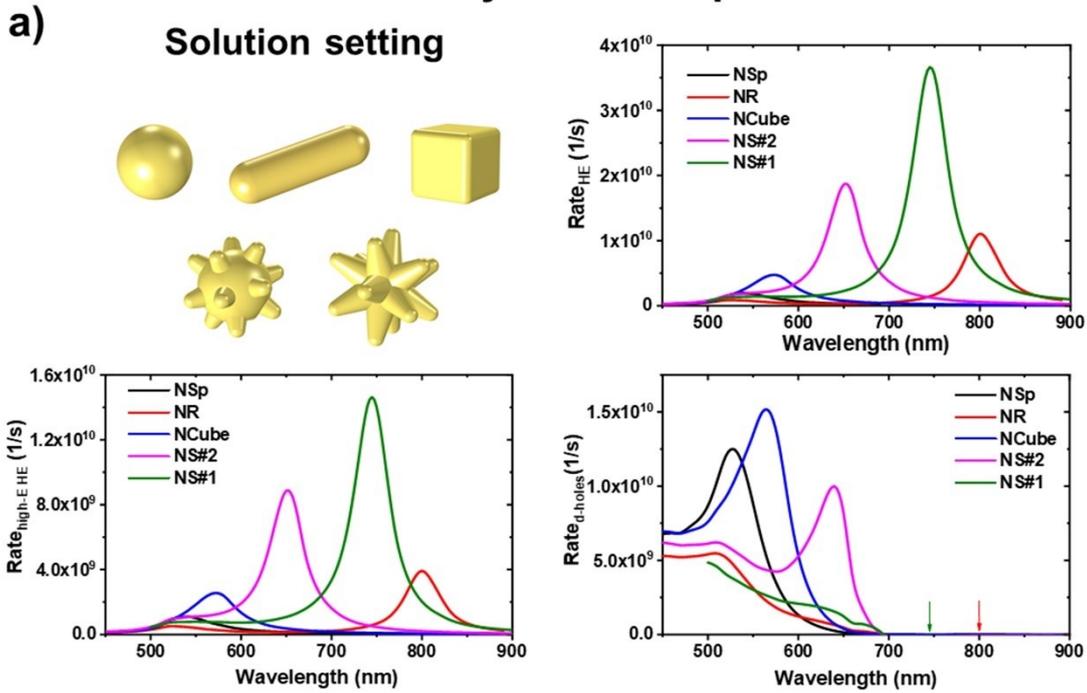

## Efficiencies

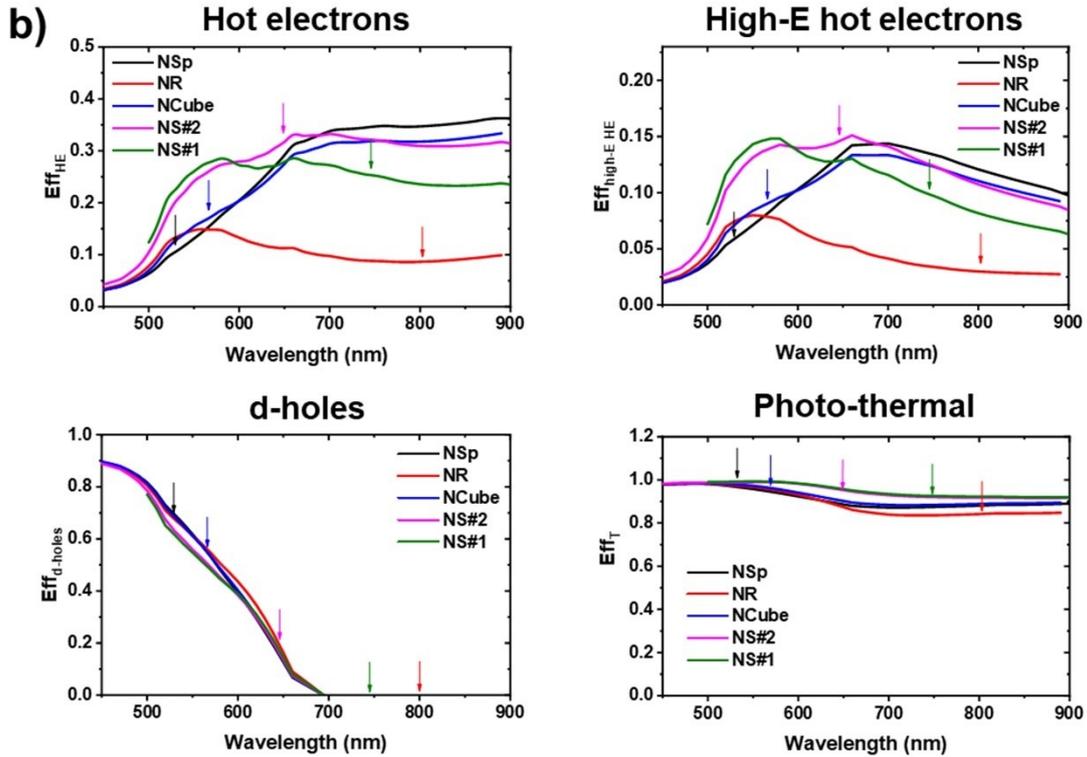

**Figure 7.** Optical and optoelectronic properties of Au NCs for non-polarized excitation. a) Generation rates for the HEs, high-E HEs, and d-holes in NCs with $R_{eff} = 15$ nm. b) Efficiencies of generation of excited carriers and heat. The vertical arrows show the plasmon peak positions of NCs on the curves of efficiencies.

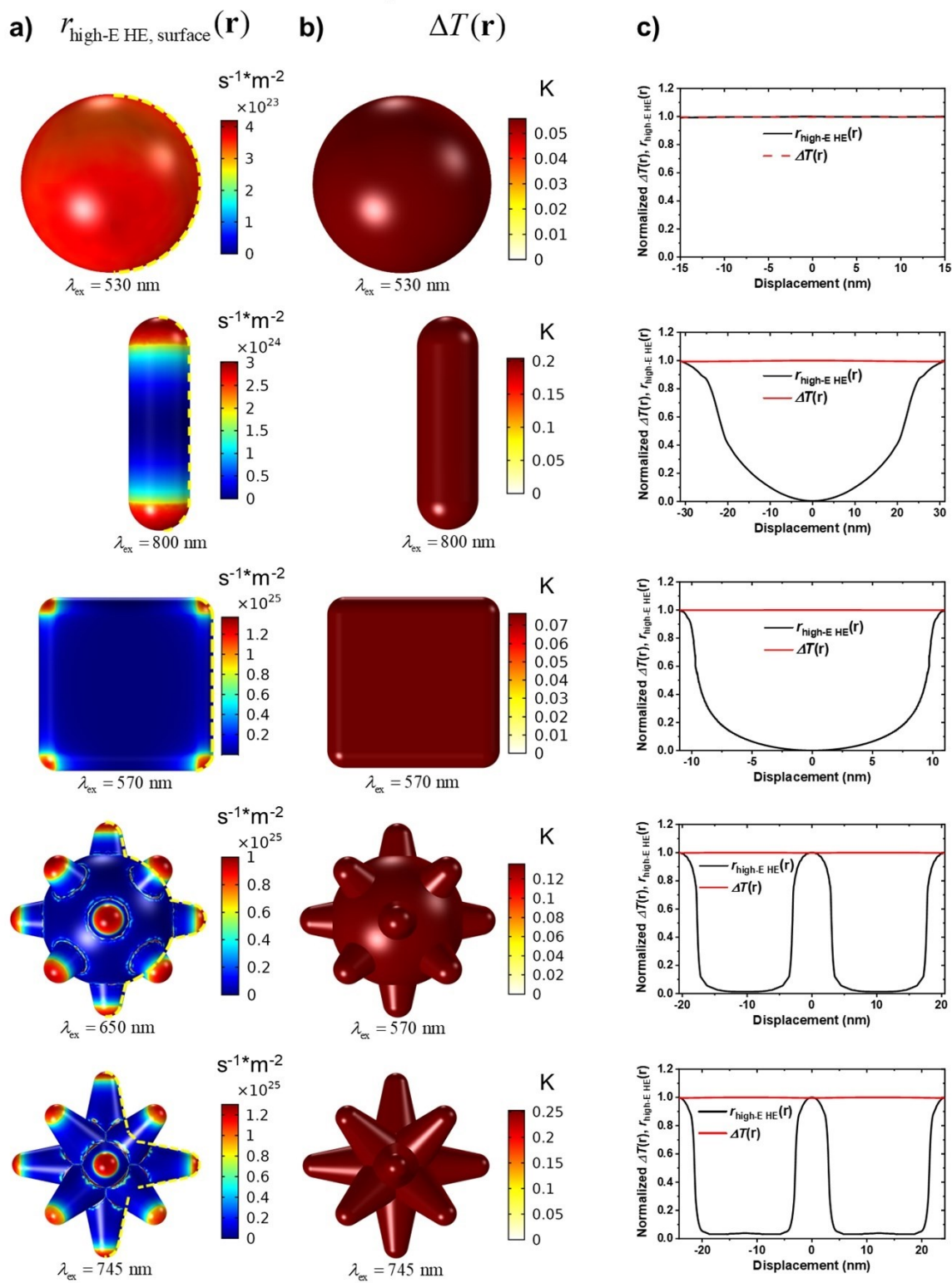

**Figure 8.** Local distributions of the rates and phototemperature surface maps on the NCs for non-polarized excitation (solution setting). The maps are averaged over the excitation

configurations, as described in the text. The excitation wavelength for the NC is taken at its plasmon peak; $R_{eff} = 15\,\text{nm}$ for all NCs. a) Surface maps for the generation rates b) Surface maps for $\Delta T$. c) The curves describe the anisotropy of the spatial distributions. The profile data were plotted along the yellow dashed lines shown in (a). The normalization is done by dividing by the maximum value. The excitation wavelengths ($\lambda_{ex}$) are taken at the plasmon peaks for all shapes.

## Solution setting and non-polarized excitation: The bulk maps of d-hole generation

a)

$\lambda_{ex} = 530$ nm

$\lambda_{ex} = 520$ nm

$\lambda_{ex} = 570$ nm

$\lambda_{ex} = 650$ nm

b)

**Figure 9.** a) Local 2D maps of the d-hole generation rate for various NCs under nonpolarized excitation. The maps are again averaged over the excitation configurations, as described in the text. The effective radius for all NCs is the same, $R_{eff} = 15\,\text{nm}$. b) The d-hole generation rate near the surface and inside, plotted along the dashed lines on panel (a); the distances between the lines and the surface are 0, 2.5, and 5 nm. The excitation wavelengths ($\lambda_{ex}$) are taken at the plasmon peaks for all shapes.

## 5. Conclusions

Currently, the field of plasmonic photocatalysis enjoys much interest and is gathering many exciting research results.[17,30,35,36,73,74,76–79] Here we presented a theoretical formalism allowing us to compute local photochemical and photophysical properties for plasmonic NCs with complex shapes. The theory is based on the numerical solution of an integral nonlinear equation for the Kreibig's plasmon-decay parameter, $\gamma_S$,[14,79] which can be achieved using a variety of standard software tools. As an example, one convenient computational tool, used here by us, is COMSOL Multiphysics. The most interesting behaviors appear, of course, for NCs with hot spots, where the local photophysical hot-carrier response becomes strongly enhanced. There is a striking difference between the local distributions of hot-carrier generation and phototemperature. Whereas the local intraband HE and d-hole rates are typically anisotropic and non-uniform in space, the local plasmonic phototemperature in small NCs is always nearly uniform, due to the metal's high thermal conductivity. Regarding the local distributions of the nonequilibrium properties, it is important to compute the setting with a substrate and the solution case separately. For example, the hot carrier's maps for small spherical NPs are uniform for the intraband HEs (solution) and interband HHs (both solution and substrate settings). In contrast, those maps are non-uniform (anisotropic) for spherical NCs in the substrate model. Furthermore, in most cases, the symmetry of the hot-carrier maps is different in the solution and substrate settings. Finally, we comment that noble metals, such as Au, Ag, and Cu, serve as model systems, but our theory applies to a wide range of conductors with plasmonic responses.[80] To give a few examples, one should mention the TiN and ZrN material systems [38,81] and various hybrid nanostructures.[82,83]

**Supporting Information**
Supporting Information is available from the Wiley Online Library or from the author.

**Acknowledgements**

A.M and Z.M.W. acknowledge the National Key Research and Development Program of China (2019YFB2203400) and the "111 Project" (B20030). E.Y.S. and A.O.G. are supported by the Nanoscale & Quantum Phenomena Institute at Ohio University. L.V.B. acknowledges support from the Xunta de Galicia (Centro singular de investigación de Galicia accreditation 2019-2022), the European Union (European Regional Development Fund - ERDF), the National Natural Science Foundation of China (Project No. 12050410252) and the Spanish Ministerio de Ciencia e Innovación under Project PID2020-118282RA-I00. This work was also funded (M.A.C.-D.) by the Ministerio de Economía y Competitividad de España (CTM2017-84050-R), Ministerio de Ciencia e Innovación (PID2020-113704RB-I00), Xunta de Galicia (Centro Singular de Investigación de Galicia - Accreditation 2019-2022 ED431G 2019/06 and IN607A 2018/5), and European Union-ERDF (Interreg V-A - Spain-Portugal 0245_IBEROS_1_E, 0712_ACUINANO_1_E, and 0624_2IQBIONEURO_6_E, and Interreg Atlantic Area NANOCULTURE 1.102.531). S. B. acknowledges funding by the Deutsche Forschungsgemeinschaft under Germany´s Excellence Strategy – The Berlin Mathematics Research Center MATH+ (EXC-2046/1, project ID: 390685689) and by the German Federal Ministry of Education and Research (BMBF Forschungscampus MODAL, project number 05M20ZBM). Finally, A.O.G. appreciates the generous support from the Berlin MATH+ Center and Zuse-Institute Berlin via the 2021 MATH+ Distinguished Visiting Scholarship Award.

**Plasmonic Nanocrystals with Complex Shapes for Photocatalysis and Photogrowth: Contrasting Anisotropic Hot-Electron Generation with the Photothermal Effect**

*Artur Movsesyan, Eva Yazmin Santiago, Sven Burger, Miguel A. Correa-Duarte, Lucas V. Besteiro,* Zhiming Wang,* and Alexander O. Govorov**

A. Movsesyan, Z. Wang,
Institute of Fundamental and Frontier Sciences, University of Electronic Science and Technology of China, Chengdu 610054, China
E-mail: zhmwang@uestc.edu.cn
A. Movsesyan, E. Y. Santiago, A. O. Govorov,
Department of Physics and Astronomy and Nanoscale and Quantum Phenomena Institute, Ohio University, Athens OH 45701, USA
E-mail: govorov@ohio.edu
S. Burger
Zuse Institute Berlin, 14195 Berlin, Germany; JCMwave GmbH, 14050 Berlin, Germany
M. A. Correa-Duarte, L.V. Besteiro,
CINBIO, Universidade de Vigo, 36310 Vigo, Spain
E-mail: lucas.v.besteiro@uvigo.es
Z. Wang
Institute for Advanced Study, Chengdu University, Chengdu 610106, China

**Table S1:** The parameters for gold, entering the Drude fit for the empirical dielectric constant[S1] in the long-wavelength limit:

$$\varepsilon_{\text{metal, Drude}}(\omega) \approx \varepsilon_{\text{b,Drude}} - \frac{\omega_p^2}{\omega(\omega + i[\gamma_D + \gamma_s])}.$$

Note that the rates and plasmon frequency are given here in energy units.

| Parameters | Au |
|---|---|
| $\gamma_p$ | 0.073 eV |
| $\hbar / \tau_{ph}$ | 0.0013 (0.5 ps) |
| $\omega_{p,\text{Drude}}$ | 9.1 eV |
| Work Function | 4.6 eV |
| Fermi Energy | 5.5 eV |
| $\varepsilon_{b,\text{Drude}}$ | 9.07 |

**Table S2:** Thermal parameters of water and gold used in the computations.

| Water | Gold |
|---|---|
| $k = 0.6$ W/m·K | $k = 318$ W/m·K |
| $c = 4181$ J/kg·K | $c = 129$ J/kg·K c |
| $\rho = 1000$ kg/m$^3$ | $\rho = 19300$ kg/m$^3$ |

1. **Numerical formalism for the quantum collisional broadening, $\gamma_s$.**

A detailed theory and all needed derivations can be found in our previous paper, Ref. [S2]. The following non-linear integral equation should be solved:

$$\gamma_s = \frac{3}{4} \cdot v_F \frac{\int_S |E_{\omega, \text{normal}}(\theta, \varphi)|^2 ds}{\int_{NC} \mathbf{E}_\omega \cdot \mathbf{E}_\omega^* dV}.$$

Here [$v_F$] $v_F$ is the Fermi velocity in a metal. A convenient method to solve it is based on COMSOL and the iterative procedure.[S2,S3] As the first step in our iteration procedure, we assume $\gamma_s \equiv 0$; then, just a few iterations will take us to a reliable number, since the method converges really fast even for a NC with a complex shape. For a spherical NC, $\gamma_s$ is given by the following simple equation:[S4]

$$\gamma_{s,sphere} = \frac{3}{4} \cdot \frac{v_F}{R_{NSp}}.$$

## 2. Discussion and estimates for the temperature formalism.

The thermal transport equation in the main text (Equation 13) provides us with a microscopic description of the phototemperature distribution. This equation should be solved numerically. However, it is possible to estimate analytically the important time scales defined in the main text ($\tau_{T,Au}/\tau_{transfer}$ and $\tau_{T,Au}/\tau_{transfer}$). For the heat diffusion time inside the metal, we just employ the standard equation of the diffusion length: $l_{diff} = \sqrt{K_{Au} t}$, where $K_{Au} = k_{Au}/(c_{Au}\rho_{Au})$ is the thermal diffusivity of gold. Then, taking $l_{diff} = R_{eff}$, we obtain the heat diffusion time in the NC:

$$\tau_{T,Au} = \frac{R_{eff}^2}{K_{Au}}.$$

Then, we are getting: $\tau_{T,Au} = 1.8\,ps$ for $R_{eff} = 15\,nm$. As expected, this time is short since gold is an excellent conductor of heat – almost "a superconductor". How to estimate the heat release time, i.e., $\tau_{T,Au}/\tau_{transfer}$? For this, one can employ the phenomenological energy-balance approach:

$$\frac{\partial \Delta U_{NC}}{\partial t} = Q_{NC}(t) - \frac{\Delta U_{NC}}{\tau_{transfer}}, \qquad (S1)$$

where $\Delta U_{NC} = \rho_w c_{Au} V_{NC} \Delta \bar{T}$ is the heat accumulated by the NC, and $Q_{NC}(t)$ is the total absorption power of a NC; $V_{NC}$ and $\Delta \bar{T}$ are the NC volume and the average temperature inside the NC, respectively. We note that the approach of Equation (S1) was widely used a lot in both experimental and theoretical papers (e.g., see Refs. [S5–S7]). Then, for the CW regime, we should have $\partial \Delta U_{NC}/\partial t = 0$ and

$$Q_{NC} - \frac{\Delta U_{NC}}{\tau_{transfer}} = 0. \qquad (S2)$$

In the steady state, the phototemperature at the surface of a spherical NC is:

$$\Delta T_{surface} = \frac{Q_{NC}}{4\pi k_{water}} \frac{1}{R_{eff}}.$$

And, using Equation (S2), and since $\Delta \bar{T} \approx \Delta T_{surface}$, we obtain the following equation:

$$\tau_{transfer} = \frac{1}{3} \rho_{Au} c_{Au} \frac{R_{eff}^2}{k_{water}} = \frac{1}{3} \frac{\rho_{Au} c_{Au}}{\rho_w c_w} \frac{R_{eff}^2}{K_{water}}. \quad (S3)$$

The above expression is very useful as an estimate. For our conditions and $R_{eff} = 15$ nm, $\tau_{transfer} = 310$ ps. The letter is consistent with the available time-resolved experimental data.[S8] As expected, we observe that $\tau_{T,Au} / \tau_{transfer} \ll 1$. Strictly speaking, Equation S3 was derived for a spherical NC; however, one can use this equation to estimate the heat release time for an arbitrary NC shape.

As we observe, NCs with complex shapes, like nanostars and nanocubes, possess a strongly-nonuniform distribution of heat generation, i.e., $q_{local}(r)$. Therefore, for large sizes, we should expect a nonuniform temperature generation in a NC. To analyze this situation, we need to look at the parameter ($\alpha$) characterizing the temperature variance ($\delta T$) inside an NC:

$$\alpha = \tau_{T,Au} / \tau_{transfer} = \frac{\frac{R_{eff}^2}{K_{Au}}}{\frac{1}{3} \frac{\rho_{Au} c_{Au}}{\rho_w c_w} \frac{R_{eff}^2}{K_{water}}} = 3 \frac{K_{water}}{K_{Au}} \frac{\rho_w c_w}{\rho_{Au} c_{Au}} \neq f(R_{eff}).$$

First, we see that this parameter does not depend on the size, and it seems that $\delta T$ should be small for a NC of arbitrary size. However, one should note that the above result is valid only under certain conditions, which are the following:

(a) An NC should be of a quasi-spheric shape, where the irregularities do not play an essential role in a spatial T-distribution.

(b) An NC should have a small size when the near-field approximation is valid. In this case, the electric field can still be strongly nonuniform due to hot spots (see our data for nanostars and nanocubes); however, the electromagnetic field is not shielded by the NC, and the field distribution does not depend essentially on the k vector.

In general, large NCs behave differently. For large sizes, the electromagnetic field will be shielded by the NC, and the energy dissipation will appear only on one side of the NC. We can see it for large spheres treated within the Mie theory.[S9–S11] Therefore, for large sizes and complex geometries with narrow spaces, one should be careful using the above estimates, which were made for small, quasi-spherical NCs. For example, in the case of a nano-shuriken in Ref. [S12], the T-distribution along the thin and long Au arms is strongly nonuniform; this property was confirmed both experimentally and theoretically. The length of the thin arms in the above shuriken structure ~ 150 nm. To conclude, large plasmonic structures with strong electrodynamic effects, complex surfaces, and narrow spaces can show strongly nonuniform T-distributions.

## 4. Additional numerical data.

Figures S1-S3 below include additional supporting data for our formalism. Figure S1 shows the consistence of our iteration formalism for the quantum parameter. Furthermore, Figures S2 and S3 show the local Drude absorption power, which reflect geometry of NCs.

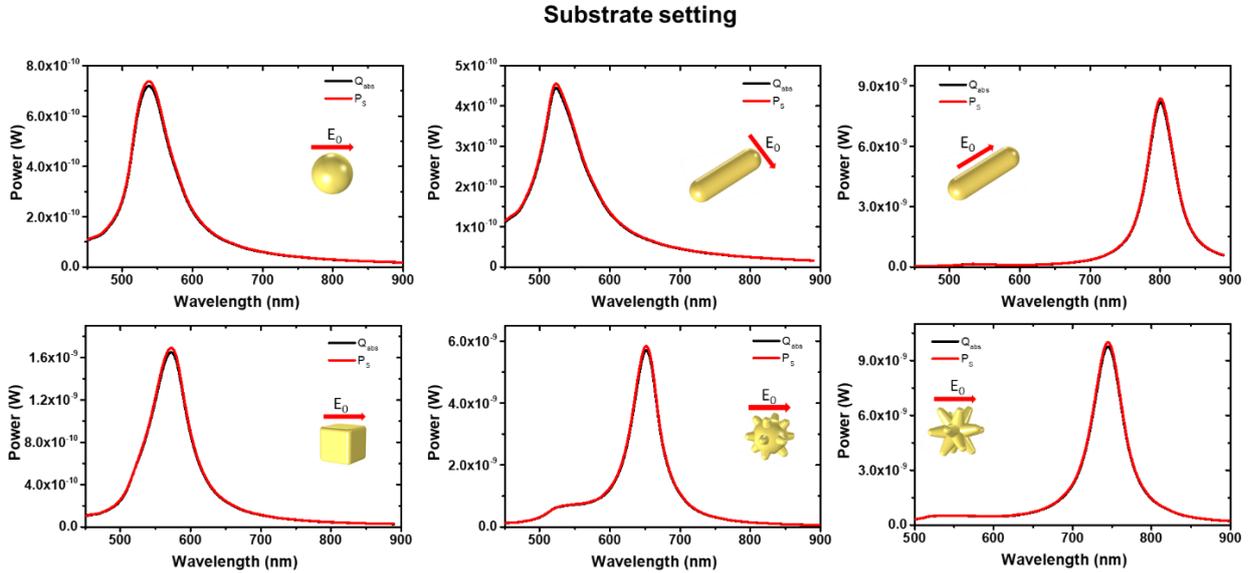

**Figure S1:** Power absorbed due to the surface-mediated scattering mechanism, computed by two different formalism – $P_s$ and $Q_{abs}$.

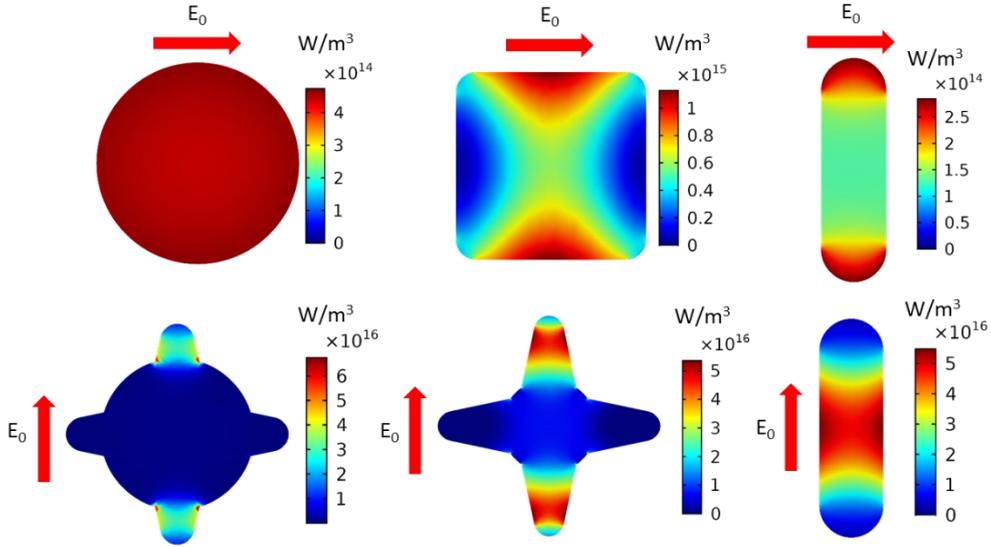

**Figure S2:** Local maps of the Drude absorption power in the NCs for polarized excitation. The polarization is oriented along the red arrow. The effective size is taken to be $R_{eff}=15\,\text{nm}$, for all NPs.

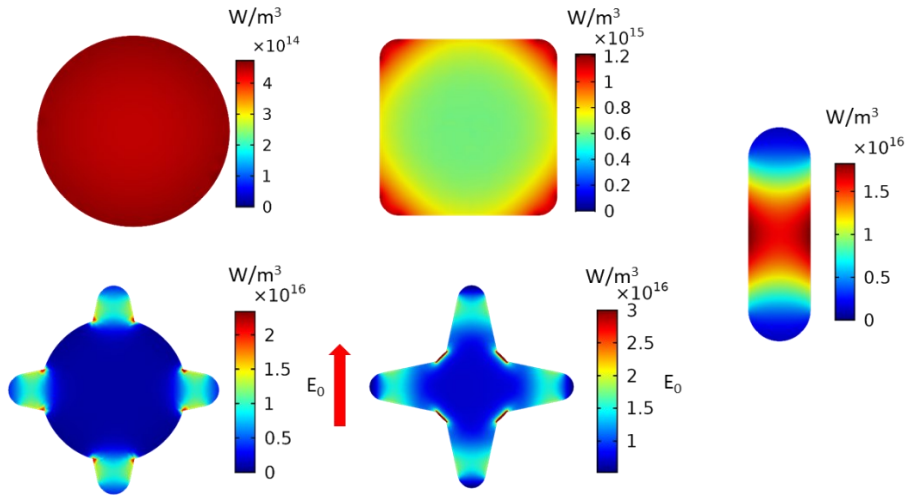

**Figure S3:** Local maps of the Drude power in our NCs for nonpolarized excitation. The maps are averaged using the 12 excitation configurations. The effective size assumed for all NCs - $R_{eff}=15\,\text{nm}$.

## 5. An Ab-initio, hybrid, and field-theoretical approaches to the plasmonic problem.

Our COMSOL-based formalism developed here utilizes Kreibig's idea of a modified local dielectric function incorporating the quantum parameter $\gamma_s$. This parameter should be carefully evaluated from the integral equation for complex geometries. Together with the Drude and d-hole generations, the surface-assisted intraband HE mechanism forms a closed set of physical parameters to describe the hot-e problem of a plasmonic NC. The justification of our approach comes directly from the equation of motion for the density matrix.[S13] Of course, there are many more approaches to this problem, and we will name here a few.

First, we should mention celebrated *ab initio* methods, DFT and TD-DFT, solving the atomistic Kohn-Sham Hamiltonians directly.[S14] The figure below shows the TD-DFT calculation of a Ag cluster exhibiting both HEs and Drude electrons from Ref [S15] (for the earlier paper by the same group, please see Ref. [S16]). The appearance of the Drude electrons indicates the formation of the plasmon wave. However, this Ag cluster is still tiny (~ few nanometers), and the Drude peaks are not overwhelming; in NCs with sizes >10nm, the amplitude of the Drude peaks at low energies in the HE distribution grows dramatically.

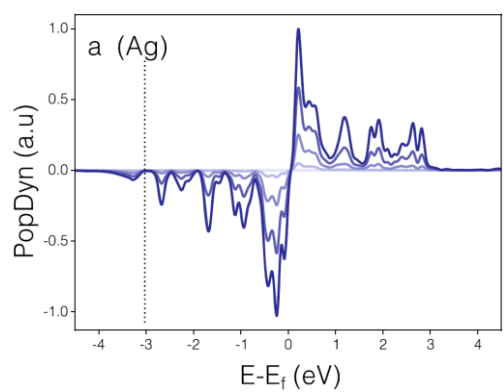

Figure S4: HE distributions from the *ab initio* TD-DFT of a Ag cluster from Ref. [S15]. Reproduced with permission.[S15] Copyright 2020, American Chemical Society.

Another stream of papers on HEs utilizes the rate derived initially from the quantum equations of motion; such a rate should incorporate a phonological relaxation time $\tau$ .[S17,S18] On the other hand, our formalism[S2, S13] is intrinsically more complex (originating from the equation of motion for the density matrix) and incorporates more relaxation parameters and also relaxation

integrals.[S2,S13] Overall, the results of Refs. [S2,S13,S17] and [S18] are qualitatively similar and agree with the *ab initio* TD-DFT picture in Ref. [S15] (Figure S4).

A field-theory approach (i.e., second quantization) for plasmonic fields was introduced a long time ago – please see a chapter in this classical text.[S19] For NCs, the second quantization (together with the Lindblad relaxation operator) was successfully used in Ref. [S20], treating nontrivial exciton-plasmon settings in the nonlinear Rabi-Fano regime. Futhermore, the field-theory approach was combined with the *ab initio* DFT-calculated electron wavefunctions in plasmonic slabs in Refs. [S21,S22]. Ref. [S21] focuses on the interband absorption in a Ag slab; this theory starts from the second quantization of surface-plasmon polaritons and the formulation of the plasmon-electron field operator. The second-quantization formalism is, of course, based on the phenomenological local dielectric function. It is a hybrid approach allowing us to see the distribution of interband electrons and holes. Ref. [S22] used a similar methodology but combined the atomistic DFT for wavefunctions with the electron-phonon field operators. Of course, the above short notes are not a comprehensive review of quantum theories of plasmon excitations and plasmonic HEs, which have a long history.